\documentstyle[12pt,epsf]{article}
\def\simge{\mathrel{%
   \rlap{\raise 0.511ex \hbox{$>$}}{\lower 0.511ex \hbox{$\sim$}}}}
\def\simle{\mathrel{
   \rlap{\raise 0.511ex \hbox{$<$}}{\lower 0.511ex \hbox{$\sim$}}}}
\def\slashchar#1{\setbox0=\hbox{$#1$}           
   \dimen0=\wd0                                 
   \setbox1=\hbox{/} \dimen1=\wd1               
   \ifdim\dimen0>\dimen1                        
      \rlap{\hbox to \dimen0{\hfil/\hfil}}      
      #1                                        
   \else                                        
      \rlap{\hbox to \dimen1{\hfil$#1$\hfil}}   
      /                                         
   \fi}                                         %

\def\dfrac#1#2{{\displaystyle {#1 \over #2}}}
\def\dsum{\mathop{\displaystyle \sum }}

\newcommand{\ewxy}[2]{\setlength{\epsfxsize}{#2}\epsfbox[45 240 320 350]{#1}}

\newcommand{\msb}{\overline{\rm{MS}}}
\newcommand{\be}{\begin{equation}}
\newcommand{\ee}{\end{equation}}
\newcommand{\bea}{\begin{eqnarray}}
\newcommand{\eea}{\end{eqnarray}}
\newcommand{\dl}{\stackrel{\leftarrow}{D}}
\newcommand{\dr}{\stackrel{\rightarrow}{D}}
\newcommand{\dcl}{\stackrel{\leftarrow}{\cal D}}
\newcommand{\dcr}{\stackrel{\rightarrow}{\cal D}}
\newcommand{\dsl}{\stackrel{\leftarrow}{\slashchar D}}
\newcommand{\dsr}{\stackrel{\rightarrow}{\slashchar D}}
\newcommand{\wcl}{\stackrel{\leftarrow}{W}}
\newcommand{\wcr}{\stackrel{\rightarrow}{W}}

\newcommand{\cals}{{\cal S}}
\newcommand{\calo}{{\cal O}}
\newcommand{\vecp}{{\vec p}}

\begin{document}
\pagestyle{empty} 
\vspace{-1.0truecm}
\begin{flushright}
CERN-TH/97-15  \\
ROME prep. 1163/97 \\
ROM2F/97/23 \\
\end{flushright}
\centerline{\LARGE{\bf Improved Renormalization of Lattice}}
\vskip 0.2cm
\centerline{\LARGE{\bf Operators: A Critical Reappraisal}}
\vskip 0.3cm
\centerline{\bf{M. Crisafulli$^{a}$, V. Lubicz$^{b}$ and
A. Vladikas$^{c,d}$}}
\vskip 0.3cm
\centerline{$^a$ Dip. di Fisica, Univ. di Roma ``La Sapienza'' and
INFN, Sezione di Roma,}
\centerline{P.le A. Moro 2, I-00185 Roma, Italy.}
\centerline{$^b$ Dip. di Fisica, Univ. di Roma Tre
and INFN, Sezione di Roma,}
\centerline{Via della Vasca Navale 84, I-00146 Roma, Italy}
\centerline{$^c$ Theory Division, CERN, 1211 Geneva 23, Switzerland.}
\centerline{$^d$ INFN, Sezione di Roma II, and 
Dip. di Fisica, Univ. di Roma ``Tor Vergata'',}
\centerline{Via della Ricerca Scientifica 1, I-00133 Roma, Italy.}

\abstract{
We systematically examine various proposals which aim at increasing the accuracy
in the determination of the renormalization of two-fermion lattice 
operators. We concentrate on three finite quantities which are particularly
suitable for our study: the renormalization constants of the vector and
axial currents and the ratio of the renormalization constants of the
scalar and pseudoscalar densities. We calculate these quantities
in boosted perturbation theory, with several running boosted couplings,
at the ``optimal" scale $q^\ast$. We find that the results of boosted
perturbation theory are usually (but not always) in better agreement  with
non-perturbative determinations of the renormalization constants than
those obtained with standard perturbation theory. The finite 
renormalization constants of two-fermion lattice operators are also obtained
non-perturbatively, using Ward Identities, both with the Wilson and the
tree-level Clover improved actions, at fixed cutoff ($\beta=6.4$ and $6.0$
respectively). In order to amplify finite cutoff effects, the quark masses 
(in lattice units) are varied in a large interval $0 \simle am \simle 1$. 
We find that discretization effects are always large with the Wilson action,
despite our relatively small value of the lattice spacing ($a^{-1} \simeq
3.7$ GeV). With the Clover action discretization errors are significantly
reduced at small quark mass, even though our lattice spacing is larger
($a^{-1} \simeq 2$ GeV). However, these errors remain substantial in the heavy
quark region. 
We have implemented a proposal for reducing $\calo (am)$ effects, which
consists in matching the lattice quantities to their continuum counterparts
in the free theory. We find that this approach still
leaves appreciable, mass dependent, discretization effects. 
}
\vskip 0.5cm
\begin{flushleft}
CERN-TH/97-15 \\
July 1997
\end{flushleft}
\vskip 0.5 cm
\vfill\eject

\pagestyle{empty}\clearpage
\setcounter{page}{1}
\pagestyle{plain}
\newpage
\pagestyle{plain} \setcounter{page}{1}

\section{Introduction}
\label{sec:intro}

One of the major goals of lattice QCD computations is the non-perturbative 
evaluation of hadronic Matrix Elements (ME's) which are relevant for strong
and electroweak interactions. On the lattice, the ME's of bare operators
are computed at finite UV cutoff $a^{-1}$ ($a$ is the lattice
spacing) by simulating QCD at fixed bare coupling $\beta = 6/g_0^2$.
These ME's are functions of the finite cutoff. In order to
extract physical information from them, the lattice operators require
renormalization. For example, the
$\rho$-meson decay constant $f_\rho^{-1}$ is obtained from the ME
$\langle 0 \vert V_k(0) \vert \rho_r \rangle$ as follows: 
\be
\epsilon_k^r \frac{m_\rho^2}{f_\rho} = Z_V
\langle 0 \vert V_k(0) \vert \rho_r \rangle
\label{eq:frho}
\ee
where $m_\rho$ is the meson mass and $\epsilon^r_k$ is its polarization
vector. $V_k$ is a spatial component of the bare (lattice) vector current 
and $Z_V$ is its renormalization constant (RC). $Z_V$ connects the lattice 
theory at finite cutoff to the renormalized continuum theory. Analogously, 
the $\pi$-meson's decay constant $f_\pi$ is determined from the ME of the
axial current through: 
\be
i f_\pi m_\pi = Z_A
\langle 0 \vert A_0(0) \vert \pi \rangle
\label{eq:fpi}
\ee
where $m_\pi$ is the pion mass, $A_0$ is the time component of the bare 
(lattice) axial current and $Z_A$ is its RC. These are specific examples in 
which operator renormalization is an essential step for passing from lattice 
to continuum physics. This renormalization is an inevitable requirement
in those cases where the ME's of bare operators diverge in the continuum limit
(e.g. the ME's of the $\Delta I = 1/2$ four-fermion operators).

So far, three methods have been implemented to the computation of the $Z$'s:
\begin{itemize}
\item{Lattice Perturbation Theory (PT) \cite{msmith}- \cite{morn}}
\item{Non-perturbative use of Ward Identities (WI's), either on hadronic states 
\cite{mm}-\cite{ukzs} or on quark states \cite{clvwi,gri,jlqcd_lat96}.}
\item{Non-perturbative (NP) renormalization on external quark and/or gluon
states \cite{np}-\cite{rossi}.}
\end{itemize}
Lattice PT can be applied to the calculation of finite and logarithmically
divergent RC's%
\footnote{PT cannot be applied to power divergent RC's, because of the
possible presence of finite non-perturbative contributions 
(see, for example, \cite{vlad,power} and references therein).}.
The WI method is only applicable to finite RC's, whereas the NP one can be 
applied to all cases. The RC's, calculated to $\calo (g_0^{2n})$ in PT,
suffer from $\calo (g_0^{2(n+1)})$ systematic errors (in most cases $n=1$).
On the other hand, non-perturbative determinations of the RC's (WI or NP method),
are subject to $\calo (a)$ systematic errors, since they are obtained from
correlation functions
computed at finite cutoff. Thus, physical quantities like those of
eqs.~(\ref{eq:frho}) and (\ref{eq:fpi}), will suffer from $\calo (g_0^{2(n+1)})$ 
and $\calo (a)$ systematic errors if the RC's are obtained in PT, and from
$\calo (a)$ errors if they are computed non perturbatively. Therefore WI or
NP estimates of the RC's are clearly preferable in principle.

In recent years, there have been considerable efforts in refining the 
perturbative computation of the RC's. Boosted PT (BPT) has been extensively 
explored by Lepage and Mackenzie \cite{lep_mac}. It consists in judicious 
choices of the expansion parameter (motivated by the behaviour of the resulting 
perturbative series, by tadpole resummation arguments and even by Mean Field 
Theory) which should bring the perturbative estimate of the RC's closer to 
their actual values. Even though the error remains $\calo (g_0^{2(n+1)})$,
any of the several choices of BPT, proposed in \cite{lep_mac},
might improve the asymptotic convergence of the perturbative series. 

On the non-perturbative side, Symanzik-type improved actions and operators have 
been constructed, in order to systematically reduce the finite cutoff error 
inherent in numerical computations. The implementation of the tree-level 
improved Clover action and operators \cite{sw,heatlie} reduces the 
discretization errors from $\calo (a)$ to $\calo(g_0^2 a)$ (more precisely, 
it removes all $\calo (a g_0^{2n} \ln^n a)$ systematic errors, which are 
effectively $\calo (a)$ in the scaling limit, $g_0^2 \sim 1/\ln a$). Recently, 
a non-perturbative extension of the tree-level Clover improvement programme
has been proposed in ref.~\cite{lusch1}, which corrects
up to $\calo(a^2)$ errors. The general proposal of these works is directly
applicable in the chiral limit (with the exception of the improvement of the
vector current which has also been implemented at non-zero quark mass; see
ref.~\cite{lusch}). Another approach to the non-perturbative improvement 
programme, which can also be applied with finite values of the quark masses,
can be found in in refs.~\cite{mrsstt,noi}.

Regarding the RC's, the main achievement of the tree-level improvement 
programme has been the reduction of discretization errors from about $20\%$
to about $5\%$ in computations with light quark masses \cite{masavla}. The 
situation is more problematic with heavy quark masses, say $am \simge 0.4$. 
Even
with a tree-level improved action, the dominant $\calo(g_0^2 am)$ error is 
appreciable. The most direct way of reducing these errors consists in reducing
the lattice spacing (by going to larger $\beta$ values), but the increase in
computational cost becomes prohibitive. Alternatively, one can try to 
reduce systematically these errors at fixed cutoff. The proposal of refs. 
\cite{lep} - \cite{ekm} is such an effort. It consists in matching lattice 
quantities to their continuum counterparts at vanishing spatial momentum,
$ap$, and large finite masses, $am$. The aim is to eventually remove all $am$ 
corrections, leaving predominantly $\calo(ap)$ errors. This program has been 
explicitly carried out only in the free theory; we refer to it as the KLM
approach. Speculations about higher order extensions can be found in
\cite{ekm}. 

This work is a thorough comparison of the various estimates (perturbative and
non- perturbative) of the RC's of two-fermion lattice operators. We have tried
to be
exhaustive in investigating the numerous improvement proposals found in the
literature. In particular:
\begin{itemize}
\item{
We have concentrated on the finite RC's of the vector and axial currents
($Z_V$ and $Z_A$) and the finite ratio $Z_S/Z_P$ of the RC's of the scalar
and pseudoscalar densities.
}\item{
We have examined the influence of boosting to the 1-loop perturbative
estimates of the above quantities. Several boosting recipes of 
ref.~\cite{lep_mac} have been applied. The ``optimal" scales of the boosted
running coupling (the $q^\ast$ of ref.~\cite{lep_mac}) has been obtained 
for $Z_V$, $Z_A$ and $Z_S/Z_P$. The $q^\ast$ values can be found in table
\ref{tab:qst}. We find that all recipes of
BPT give the same results within a few per cent ($10\%$ in the worst case).
These results are summarized in table \ref{tab:bpt}. 
Moreover, all BPT results differ from those obtained in SPT by up to $20\%$.
When compared with the RC's obtained from WI's (see table \ref{tab:zz}), BPT
results are usually in better agreement than the SPT ones. 
}\item{
The RC's $Z_V$, $Z_A$ and the ratio $Z_S/Z_P$
have subsequently been computed non- perturbatively,
using the WI method, for several quark masses. Both the Wilson and the
tree-level Clover actions have been implemented. Discretization errors
at small quark mass, being $\calo(a)$ tend to be about $20\%$ - $30\%$
in the Wilson case. In the tree-level Clover case, being $\calo(g_0^2 a)$,
these errors drop below $5\%$. However, for masses $ma \simge 0.4$,
which are used in current lattice QCD simulations, they grow beyond $10\%$.
By extrapolating our data in the quark mass, we obtain
extremely consistent Clover results for $Z_V$ and $Z_A$ at zero quark 
mass, shown in table \ref{tab:zz}. We have also checked that it is not
possible to optimize these results by implementing different choices of
improved operators, based on a different arbitrary admixture of the so-called
$\slashchar{D}$ and $m_0$ improving field rotations (see Subsection 
\ref{sec:clover} for the definitions of these field rotations).
}\item{
In order to control discretization errors of increasing quark mass, we have
applied the tree-level KLM matching prescription of refs.~\cite{lep} -
\cite{ekm}. In the Wilson case, the large systematic error due
to $\calo(am)$ effects is significantly reduced in some cases (see figure
\ref{fig:fig1}). There are cases, however, for which the KLM prescription
fails to correct large $\calo(am)$ effects (see figure \ref{fig:fig2}). 
Moreover, a similar analysis, applied to the Clover case, revealed that
KLM corrections fail to improve the situation. This is because the Clover
correlation functions we have examined are characterized by large 
$\calo (g_0^2am)$ errors in the region of heavy quark masses. Thus, the
KLM corrections, being of $\calo (a^2)$, cannot absorb these discretization
errors.
}
\end{itemize}

The present work is intended as a thorough
comparison of the various improvement schemes which have been widely
used in present-day, state-of-the-art lattice QCD phenomenology studies.
As such, it has not dealt with the more recent non-perturbative Clover
improvement of \cite{lusch1,mrsstt,noi}, which appears to be the most
promising way to reduce finite cutoff effects. 

The paper is organized as follows. In sect.\ref{sec:defs} we give some
basic definitions. In sect. \ref{sec:pt} we review the perturbative
calculation of the RC's of two-fermion lattice operators. In subsec.
\ref{sec:bpt} we discuss BPT and compare results for the RC's obtained from
various boosting recipes. We also present the ``optimal scales" $q^\ast$
of \cite{lep_mac} for the operators of interest (some technical subtleties
of this calculation can be found in the Appendix). In sect.\ref{sec:gen} we
review the well known vector and axial WI's on the lattice, with particular
emphasis on the conditions under which they can be used to obtain the RC's.
We have gathered in this section all WI's which are subsequently used
in sects. \ref{sec:wi} and \ref{sec:npres}, where we discuss the
practical implementation of the WI determination of RC's and present our
results for $Z_V$, $Z_A$ and $Z_S/Z_P$. In sect. \ref{sec:KLM} we review
the tree-level factors of the KLM-procedure for the Wilson correlation
functions and derive their Clover counterparts. Finally, in sect.
\ref{sec:conc} we present our conclusion.

\section{Definitions and Generalities on RC's}
\label{sec:defs}

In this section we define the quantities of interest to this paper. Besides
fixing  our notation, we also review some general properties of the RC's of 
lattice operators, with particular attention to their dependence on the
theory's parameters and renormalization scale.

We consider bilinear quark operators of the form:
\be
O^a_\Gamma(x) = \bar \psi(x) \Gamma \frac{\lambda^a}{2} \psi(x)
\label{eq:o}
\ee
where $\psi (x)$ is the quark field, $\Gamma$ denotes a generic Dirac matrix and 
$\lambda^a/2$ are the generators of the $SU(N_f)$ flavour group in the 
fundamental representation ($a = 1,\dots,N_f^2-1$). These generators satisfy: 
\bea
&& Tr[\lambda^a \lambda^b ] = 2 \delta^{ab} \nonumber \\
&& \left[ \frac{\lambda^f}{2}, \frac{\lambda^g}{2}\right]
=i f^{fgh}\frac{\lambda^h}{2} \nonumber \\
&& \left\{\frac{\lambda^f}{2}, \frac{\lambda^g}{2}\right\}
=d^{fgh}\frac{\lambda^h}{2}+\frac{\delta^{fg}}{N_f}\,I
\label{eq:ds}
\eea
where $I$ represents the identity. Specific bilinear operators will be denoted 
according to their Lorentz group transformations: the scalar and 
pseudoscalar densities are denoted by $S^a(x) = \bar \psi(x) \dfrac{\lambda^a}
{2} \psi(x)$ and $P^a(x) = \bar \psi(x) \dfrac{\lambda^a}{2} \gamma_5 \psi(x)$ 
respectively, whereas the local vector and axial vector currents by
$V^a_\mu(x) = \bar \psi(x) \dfrac{\lambda^a}{2} \gamma_\mu \psi(x)$ and
$A^a_\mu(x) = \bar \psi(x) \dfrac{\lambda^a}{2} \gamma_\mu \gamma_5 \psi (x)$.
Only such non-singlet bilinear quark operators will be considered in
this paper. In the case of the vector and axial currents, we will also
be using their point-split discretizations.

We will deal with operator insertions in $n$-point
Green's functions, with the operator usually inserted at the origin.
For example the insertion of $O_\Gamma$ in the 2-point Green's
function is
\be
G_O(x_1,x_2) = \langle \psi(x_1) O^a_\Gamma(0) \bar \psi(x_2) \rangle
\label{eq:gos}
\ee
In momentum space the insertion becomes
\be
G_O(p_1,p_2) = \int dx_1 dx_2 \exp(ip_1x_1) \exp(-ip_2x_2) G_O(x_1,x_2)
\label{eq:gp}
\ee
and the corresponding amputated Green's function is given by
\be
\Lambda_O(p_1,p_2) = \cals ^{-1}(p_1) G_O(p_1,p_2) \cals ^{-1}(p_2)
\label{eq:amp}
\ee
where $\cals (x_1 - x_2) =  \langle \psi(x_1) \bar \psi(x_2) \rangle$
is the quark propagator and $\cals (p)$ its momentum space counterpart.
We will also be using Green's functions between external hadronic states. 
These are defined in an analogous fashion. For example, the 3-point correlation 
function $\langle P(x_1) V_\mu(0) P^\dagger (x_2) \rangle$ is a 
correlation function of three bilinear operators. From its Fourier 
transform in 3-dimensional space, and with the aid of the spectral
representation formula at large euclidean time separations, the on-shell
hadronic ME  $\langle \pi ({\vec p}_1) \vert V_\mu \vert \pi ({\vec p}_2)
\rangle$ can be readily extracted.

Everything defined so far is a bare quantity, which we assume to be defined
in the lattice regularization%
\footnote{This means that the integrals of eq.~(\ref{eq:gp}) are really sums
($a^8 \sum_{x_1,x_2}$) which run over all lattice sites, labelled by $x_1$,
$x_2$, etc. The use of integrals instead of sums and Dirac functions instead
of Kronecker symbols is a question of notational convenience, which
should be transparent.}.
We opt for the lattice regularization scheme proposed by Wilson which
consists of a gluonic action \cite{wils1}, 
\be
S_g =  - \frac{6}{g_0^2}\sum_{P} Tr \bigl[ U_P + U_P^{\dagger} \bigr]
\label{eq:sg}
\ee
and a fermionic part \cite{wils2}, 
\bea
S_f & = & -a^4 \sum_{x,\mu} \frac{1}{2a} 
 \bigl[ \bar \psi (x) (1 - \gamma_\mu) U_\mu (x) \psi (x+\mu)
 + \bar \psi (x+\mu) (1 + \gamma_\mu) U^\dagger_\mu (x) \psi (x) \bigr]
\nonumber \\
&& + a^4 \sum_x \bar \psi (x) (M_0 + \frac{4}{a}) \psi (x) 
\label{eq:sf}
\eea 
where, in standard notation, $U_P$ is the Wilson plaquette,
$U_\mu(x)$ is the lattice gauge link and $g_0$ the bare coupling
constant. The fermion field is a vector in flavour space. Its 
components will be denoted by $u$, $d$, $s$ etc, or by $\psi_1$, $\psi_2$, 
$\psi_3$, etc. according to notational convenience. The diagonal bare mass
matrix  is denoted by $M_0$ and its elements by $m_{01}$, $m_{02}$, $m_{03}$
etc. (or $m_{0u}$, $m_{0d}$, $m_{0s}$ etc.). We use $m_0$ 
in the mass degenerate case. As is well known, for Wilson fermions, the 
quark mass, besides a multiplicative renormalization, is also subject to an 
additive one. Defining $m = m_0 - m_C$, the renormalized mass will be given by:
\be
m_R = Z_m m = Z_m [ m_0 - m_C ]
\label{eq:mrzm}
\ee
with $Z_m$ the multiplicative RC. The chiral limit is then $m \rightarrow 0$; i.e. 
$m_0 \rightarrow m_C$. At tree-level, $m_C = -4/a$. For the diagonal elements of 
the matrix $M_0 - m_C$ we will use either $m_1$, $m_2$, $m_3$ etc., or $m_u$, 
$m_d$, $m_s$ etc., according to notational convenience.

All operators studied in this work are subject
to multiplicative renormalization; cases involving mixing of equal-
or lower-dimensional operators (e.g. the four-fermion operators driving the
$\Delta S = 2$ or $\Delta I = 1/2$
transitions) will not be considered. Thus, the renormalized operator is
given by
\be
\hat O_\Gamma(g_R^2,m_R,\mu) = \lim_{a \rightarrow 0}
[Z_O(g_0^2,a,m,\mu) O_\Gamma(g_0^2,m,a)]
\label{eq:zodef}
\ee
where $g_R$ is the renormalized gauge coupling and $\mu$ the renormalization 
scale. Note that in the bare quantities, the bare mass $m_0$ has been traded off 
for the more convenient (from the point of view of the chiral limit) subtracted 
mass $m$. A specific, if arbitrary, choice of renormalization scheme and scale 
are implied. Suppressing the dependence on couplings, masses, scales and cutoffs 
for notational simplicity, we list other relationships of interest between bare 
and renormalized quantities (denoted in general by ``hats"): the renormalized 
quark propagator is
\be
\hat \cals (p) = \lim_{a \rightarrow 0}[Z_\psi \cals (p)]
\label{eq:cp}
\ee
where $Z_\psi^{1/2}$ is the quark field RC. From the above definitions
of $Z_O$ and $Z_\psi$, with the aid of eqs.~(\ref{eq:gos}) - (\ref{eq:amp}),
we find for the renormalized Green's functions:
\bea
&& \hat G_O(p_1,p_2) = \lim_{a \rightarrow 0}[Z_\psi Z_O G_O (p_1,p_2)] 
\nonumber \\
&& \hat \Lambda_O (p_1,p_2) = \lim_{a \rightarrow 0}[Z_\psi^{-1} Z_O \Lambda_O 
(p_1,p_2)]
\label{eq:cl}
\eea

Let us now examine the dependence of the RC's $Z_O$ on the bare parameters of 
the action. Multiplicative $Z_O$'s are dimensionless quantities. Consequently, 
their functional dependence on the bare parameters and the cutoff can in principle 
be only of the form $Z_O(g_0^2,am,a\mu)$. For the specific case of 
finite RC's (e.g. those of the vector and axial 
currents) there is no dependence on the renormalization scale $\mu$. In other
words, the anomalous dimension is zero; this will be discussed in detail in 
the following section. Therefore, we are left with a possible dependence of
the form
$Z_O(g_0^2,am)$. However, there can be no dependence of $Z_O$ on $am$. Regular 
terms such as positive powers of $am$, which may appear in the calculation of 
the bare Green's function $\Lambda_O$, will drop out of $Z_O$, since RC's are 
strictly defined in the continuum limit $a \rightarrow 0$; c.f. eq.~(\ref{eq:cl}). 
This leaves us with singular terms, such as $\ln(am)$ and negative powers of $am$. 
But since we require the existence of the chiral limit ($m \rightarrow 0$) in
QCD, such a dependence must also be ruled out. This exhausts
all possibilities, and renders $Z_O$ independent of $am$. In conclusion, the
finite RC's $Z_V$, $Z_A$ and the ratio $Z_S/Z_P$, examined in this paper,
are of the form $Z_O(g_0^2)$.%
\footnote{Dimensionless RC's which are not finite
have the general dependence $Z_O(g_0^2,am,a\mu)$. Provided that the 
renormalization scale $\mu$ is chosen to be greater then the quark masses 
($\mu \gg m$), there can be no dependence on $am$ (same argument as above). 
The requirement of a well defined chiral limit allows only regular terms in 
$m/\mu$, which however have been assumed negligible. Thus, in this case the RC's 
have the form $Z_O(g_0^2,a\mu)$}

\section{Evaluation of RC's from Perturbation Theory}
\label{sec:pt}

In this section we review the basic concepts of the perturbative evaluation of
the RC's of lattice operators and examine the proposals of ref.~\cite{lep_mac}
which aims at the improvement of the convergence of the perturbative series in
lattice PT.

The standard perturbative determination of the RC $Z_O$ of a bilinear quark
operator $O_\Gamma$ (flavour indices are suppressed, unless necessary) 
requires the calculation of the quark self- energy and the
amputated Green function $\Lambda_O(p)$ of eq.~(\ref{eq:amp}) between
external quark states, both at momentum  $p_\mu$ (see, for example, \cite{mz}).
Here we will work with the projected quark propagator $\Gamma_{\cal{S}}(p)$, 
and projected amputated Green function $\Gamma_O(p)$, defined as
\bea
&& \Gamma_\Sigma(p) = \frac{-i}{48}
Tr \left[ \gamma_\mu \frac{\partial \cals (p)^{-1}}{\partial p_\mu} \right]
\nonumber \\
&& \Gamma_O(p) = \frac{1}{12} Tr\left[P_O \Lambda_O(p)\right]
\label{eq:proj_GF}
\eea
where the trace is over spin and colour indices
and $P_O$ is the Dirac matrix
which renders the tree-level value of $\Gamma_O(p)$ equal to unity (i.e.
it projects out the nominal Dirac structure of the Green function
$\Lambda_O(p)$):
\bea
&& P_S = I \qquad ; \qquad
P_P = \gamma_5 \nonumber \\
&& P_V = \frac{1}{4} \gamma_\mu \qquad ; \qquad
P_A = \frac{1}{4} \gamma_5 \gamma_\mu
\label{eq:proj}
\eea
From eq.~(\ref{eq:cp}) we see that $\Gamma_\Sigma (p)$ renormalizes like
\be
\hat \Gamma_\Sigma (p) = \lim \left[ Z_\psi^{-1} \Gamma_\Sigma (p) \right]
\label{eq:rensig}
\ee
The renormalization of $\Gamma_O(p)$ is identical to that of
$\Lambda_O(p)$; c.f. eq(\ref{eq:cl}):
\be
\hat \Gamma_O (p) = \lim \left[ Z_\psi^{-1} Z_O \Lambda_O (p) \right]
\label{eq:renG}
\ee
In the last two equations, the limit corresponds to the removal of the
UV cutoff (e.g. $a \rightarrow 0$ on the lattice).
The choice of $\Gamma_\Sigma (p)$ and $\Gamma_O(p)$, being scalar
quantities, simplifies the discussion. Projected Green's
functions have been implemented in recent works (e.g. ref.~\cite{np}).
Another simplification consists in working with zero quark mass.
This is justified, as we have shown in the previous section, by the
lack of mass dependence of $Z_O$, in the limit $\mu \gg m$.

Upon passing from formal definitions of renormalized quantities (like
eq.~(\ref{eq:renG})) to their specific evaluation, we need to implement a
regularization. The standard practice in continuum calculations consists in
choosing a Dimensional Regularization ($DR$) scheme, such as Naive Dimensional
Regularization ($NDR$), 't Hooft - Veltman $(HV)$ etc. The expression for
$\Gamma_O$, evaluated at 1-loop PT,  can be written in the general form:
\be
\Gamma_O^{DR}(p,g_0,\epsilon)
=\left [1+\frac{g_0(\mu)^2}{(4\pi)^2} \bigl( \gamma_\Gamma
\frac{1}{\hat\epsilon} 
- \gamma_\Gamma \ln(p/\mu)^2 + C^{DR}_\Gamma \bigr) \right] + \calo(\epsilon)
\label{eq:dr}
\ee
where $\gamma_\Gamma$ is an anomalous dimension and $C^{DR}_\Gamma$ is a 
finite constant, which depends on the regularization scheme and the gauge
chosen. The factor $1 / \hat\epsilon$ in eq.~(\ref{eq:dr}) is an abbreviation
for
\be
\frac{1}{\hat\epsilon} = \frac{1}{\epsilon} + \ln(4\pi) - \gamma_E
\ee
where $\epsilon=(4-D)/2$ and $\gamma_E$ stands for Euler's constant. In $DR$
we work in $D$ dimensions, where the original bare coupling $g_0$ has
dimension $\epsilon$. Here the scale $\mu$ is introduced to render the bare
coupling $g_0(\mu)$ dimensionless. The $\mu$-dependence of the r.h.s. of
eq.~(\ref{eq:dr}) is apparent.

In order to renormalize $\Gamma_O$, we are in principle free to impose
any renormalization scheme. The $\msb$ scheme, at 1-loop, amounts to removing
the $1/ \hat\epsilon$ divergence. Since the RC of the projected Green function 
$\Gamma_O$ is given by ${Z_\Gamma} = Z_\psi^{-1} Z_O$ (see
eq.~(\ref{eq:renG})), this implies for ${Z_\Gamma}$ the value
\be
Z_\Gamma^{\msb,DR} (g_0(\mu),\epsilon) = 1 - 
\frac{g_0^2(\mu)}{(4\pi)^2} \gamma_\Gamma \frac{1}{\hat\epsilon}
\label{eq:rmsb}
\ee
Consequently, the renormalized Green function is given by
\bea
& & \hat \Gamma_O^{\msb} (p,g_{\msb}(\mu),\mu) =
\lim_{\epsilon \rightarrow 0} \left[Z_\Gamma^{\msb,DR}
(g_0(\mu),\epsilon) \Gamma_O^{DR} (p,g_0,\epsilon)\right] =
\nonumber \\
& & \qquad =  \left[1+\frac{g_{\msb}^2(\mu)}{(4\pi)^2}
\bigl( - \gamma_\Gamma \ln(p/\mu)^2 + C^{DR}_\Gamma \bigr) \right]
\label{eq:gren_msb}
\eea
where, to this order, we are free to replace $g_0$ by
$g_{\msb}(\mu)$, the $\msb$ renormalized coupling constant.

The same calculation can be repeated on the lattice. Now the UV cutoff
is provided by the inverse finite lattice spacing $a^{-1}$ and thus
the 1-loop calculation yields:
\be
\Gamma_O^{LAT}(p,g_0(a),a)=
\left[1+\frac{g_0(a)^2}{(4\pi)^2}(-\gamma_\Gamma \ln(pa)^2+C^{LAT}_\Gamma)
\right]
+\calo(a)
\label{eq:lat}
\ee
where $g_0$ is the bare coupling of the lattice action. Note that at 1-loop
the anomalous dimension $\gamma_\Gamma$ is regularization independent.

The renormalization scheme can again be chosen at will; often the
$\msb$ is chosen also on the lattice. This seemingly unnatural
choice (the $\msb$ is closely linked to continuum DR) has a few
advantages. For example, ME's of effective Hamiltonians, once calculated
non-perturbatively on the lattice, must be renormalized and combined
with perturbatively calculated Wilson coefficients, in order to obtain
physical amplitudes. The renormalization-group invariance of these
amplitudes is guaranteed only if the Wilson coefficients and the
RC's are calculated in the same renormalization scheme. 
Since the former are often known in the $\msb$ scheme,
this scheme is also preferred for the calculation of the latter.
This choice then corresponds to the requirement
\be
\hat \Gamma_O^{\msb} (p,g_{\msb}(\mu),\mu) = \lim_{a \rightarrow 0} \left[
Z_\Gamma^{\msb,LAT}(\mu a, g_0(a)) \Gamma_O^{LAT} (p,g_0(a),a)\right]
\label{eq:rgf_l}
\ee
where again the lattice coupling $g_0(a)$ should be traded for the $\msb$
renormalized coupling constant $g_{\msb}(\mu)$. This point of principle is of
limited relevance for a 1-loop calculation.
From eqs.~(\ref{eq:gren_msb}), (\ref{eq:lat}) and (\ref{eq:rgf_l}), the 
following RC is obtained:
\be
Z_\Gamma^{\msb,LAT} (\mu a,g_0(a)) = 
1 + \frac{g_0(a)^2}{(4\pi)^2} \left[ \gamma_\Gamma \ln(\mu a)^2
+C^{DR}_\Gamma - C^{LAT}_\Gamma \right]
\label{eq:rlat}
\ee
We now recall that the RC of the amputated vertex $\Gamma_O$ is 
$Z_\Gamma = Z_\psi^{-1} Z_O$. The RC of the quark field $Z_\psi$ can 
be calculated, from the renormalization of the quark propagator $\cals (p)$,
with an analogous procedure (c.f. eq.~(\ref{eq:rensig})). The result is
\be
Z_\psi^{\msb,LAT} (\mu a,g_0(a)) = 
1 + \frac{g_0(a)^2}{(4\pi)^2} \left[ \gamma_{\Sigma} \ln(\mu a)^2
+ C^{DR}_\Sigma - C^{LAT}_\Sigma \right]
\label{eq:rlatSig}
\ee
Combining eqs.~(\ref{eq:rlat}) and (\ref{eq:rlatSig}) we obtain
\be
Z_O^{\msb,LAT} (\mu a,g_0(a)) = 
1 + \frac{g_0(a)^2}{(4\pi)^2} \left[ \gamma_O \ln(\mu a)^2
+\Delta_\Gamma + \Delta_\Sigma \right]
\label{eq:rlatZ}
\ee
where
\bea
\gamma_O &=& \gamma_\Gamma + \gamma_\Sigma \nonumber \\
\Delta_\Gamma &=& C^{DR}_\Gamma - C^{LAT}_\Gamma \\
\Delta_\Sigma &=& C^{DR}_\Sigma - C^{LAT}_\Sigma \nonumber
\eea
It is this RC (with this choice of renormalization condition)
which is usually denoted by $Z_O$ in lattice PT calculations. 
The dependence of $Z_O^{\msb,LAT}$ on the coefficients
$C^{DR}_\Gamma$ and $C^{DR}_\Sigma$ comes from the choice of the $\msb$
renormalization condition (see eqs.~(\ref{eq:gren_msb}) and (\ref{eq:rgf_l}))
whereas its dependence on $C^{LAT}_\Gamma$ and $C^{LAT}_\Sigma$ from 
the lattice regularization (see eq.~(\ref{eq:lat})). Two perturbative
calculations are thus necessary, one in the continuum for the $C^{DR}$'s and
one on the lattice for $C^{LAT}$'s. The presence of the $C^{DR}$'s on the r.h.s.
of eq.~(\ref{eq:rlat}) is sometimes called regularization dependence of 
the renormalization scheme \cite{allmass,ciuc}. The reader is referred to
ref.~\cite{mz} for the explicit calculation of the RC's of several bilinear
operators with the Wilson action. The corresponding results for the Clover
action appeared in \cite{gabri,ari}.

The PT method is applicable to finite and logarithmically
divergent operators, the renormalization of which
does not require a subtraction of lower dimensional operators.
Model examples of operators with logarithmically divergent
multiplicative RC's are the scalar density $S(x) = \bar \psi_1(x) \psi_2(x)$
and the pseudoscalar density $P(x) = \bar \psi_1(x) \gamma_5 \psi_2(x)$.
These are the cases best suited to our general discussion so far.

The RC's of the vector current $V^a_\mu(x)$ and the axial current 
$A^a_\mu(x)$ are also calculable in PT on similar lines. Being (partially)
conserved quantities however, they have some peculiarities. Their (partial)
conservation guarantees zero anomalous dimension (i.e $\gamma_V = 0$ and
$\gamma_A = 0$). This result is regularization independent (e.g. valid both in
the continuum ($DR$) and the lattice ($LAT$). It implies that the current
renormalization constants are finite, scale independent quantities.
Moreover, the following useful properties must be taken into consideration:
\begin{itemize}
\item{The vector current conservation survives the continuum DR.
Thus it satisfies a WI which implies that $Z_V^{\msb,DR}=1$. This is 
an exact result, valid at all orders of PT. Therefore, at 1-loop
$V^a_\mu(x)$ has zero finite constant; i.e. $C^{DR}_V + C^{DR}_\Sigma= 0$.}
\item{As we will see in the next section, the local lattice vector current
is not a conserved quantity. It has a non-vanishing 1-loop contribution
(i.e. $C^{LAT}_V + C^{LAT}_\Sigma \ne 0$). In other words,
$Z_V^{\msb,LAT}(g_0^2) \ne 1$.}
\item{We will also show in the next section that on the lattice one
can define an extended conserved vector current, ${\widetilde V}^a_\mu(x)$,
which is also trivially renormalized with $Z_{\widetilde V}^{\msb,LAT} = 1$.
Thus, also for this current we have  $C^{LAT}_{\tilde V} + C^{LAT}_\Sigma= 0$.}
\item{For the axial current the argument is almost the same. However, some
further peculiarities arise due to the breaking of the chiral symmetry by the
regularization.
In the continuum, the finite 1-loop contribution to the renormalization
constant may or may not vanish, depending on the detailed choice of the
regularization scheme. Thus, for $NDR$ we have $C^{NDR}_A + C^{NDR}_\Sigma =
0$, implying $Z_A^{\msb,NDR}=1$. On the other hand, with $HV$ for example,
we have $C^{HV}_A + C^{HV}_\Sigma \ne 0$, implying $Z_A^{\msb,HV}(g_0^2)\ne 1$.
In the present work, we will always imply $NDR$ when referring to $DR$.}
\item{We will show in the next section that the lattice regularization
breaks chiral symmetry. Thus on the lattice $C^{LAT}_A + C^{LAT}_\Sigma \ne 0$
(for any definition of the bare lattice current, local or extended). In other
words, $Z_A^{\msb,LAT}(g_0^2) \ne 1$}
\end{itemize}

We are also interested in the finite ratio $Z_S/Z_P$. Its finiteness relies
on WI's (see next section) and implies that $\gamma_S = \gamma_P$. With the
specific choice of $NDR$ as the continuum dimensional regularization,
we also have $C^{NDR}_S = C^{NDR}_P$.

\subsection{Boosted Improvement of the PT Estimates of RC's}
\label{sec:bpt}

The perturbative calculation outlined above is only approximate, because of
the truncation of the perturbative series. At present, most 
lattice RC's are only known in PT at 1-loop. Thus, they
suffer from $\calo (g_0^4)$ systematic errors.  These errors must be born in mind
when comparing the RC's, calculated in PT, to their non-perturbative estimates. 
In this subsection we discuss these systematic errors and comment on the proposals
for improving the convergence of the perturbative series of ref.~\cite{lep_mac}.
These proposals are known as Boosted Perturbation Theory (BPT). 

We have already 
pointed out that, in principle, at 1-loop there is an ambiguity in the choice of 
couplings ($g_0(a)$, $g_0(\mu)$, $g_{\msb}(\mu)$) to be used in eq.~(\ref{eq:rlat}). 
This ambiguity is lifted in (and up to) 2-loop PT. At 1-loop 
however, a judicious choice of expansion parameter may be important in practice. 
The lattice coupling $g_0^2(a)$ has been shown to be a bad expansion parameter in 
ref.~\cite{lep_mac}, where several improved (boosted) couplings have been proposed, 
and several observables have been calculated in PT with these couplings. 

In the specific case of $Z_O$, the Standard Perturbation Theory (SPT) result is 
given as a series in the bare (lattice) coupling $g_0$
\be
Z_O(a\mu,g^2_0(a)) = 1 + C_1(a\mu) g^2_0(a) + C_2(a\mu) g^4_0(a) + \dots
\label{eq:zunb}
\ee
To first order, the above equation reduces to eq.~(\ref{eq:rlatZ}), with
$C_1(a\mu) = [\gamma_O \ln(a\mu)^2 + \Delta_\Gamma + \Delta_\Sigma ]/
(4\pi)^2$. According
to \cite{lep_mac}, perturbative series have a better behaviour if expressed
in terms of a renormalized coupling $g_V^2(q^\ast)$ at a suitable
scale $q^\ast$ (to be specified shortly):
\be
Z_O(a\mu,g^2_V(q^\ast)) = 1 + C_1(a\mu) g^2_V(q^\ast) + 
{\widetilde C}_2(a\mu,aq^\ast) g^4_V(q^\ast) + \dots
\label{eq:zb}
\ee
The claim is that the perturbative estimate obtained from the boosted series,
truncated at low orders, is closer to the non-perturbative result than its SPT 
counterpart.

The renormalized coupling $g_V(q)$ in eq.~(\ref{eq:zb}) can be chosen at will.
In ref.~\cite{lep_mac} it is suggested to define it
from the heavy quark potential $V(q)$ at momentum transfer $q$:
\be
V(q) = - \frac{4 g^2_V(q)}{3 q^2}
\label{eq:vq}
\ee
Moreover, according to \cite{lep_mac} and \cite{brodski}, the scale $q^\ast$,
which is most  appropriate to the expansion (\ref{eq:zb}) of $Z_O$, is fixed by:
\be
\ln({q^\ast}^2) = \frac{\int d^4q f(q,a\mu) \ln(q^2)}
{\int d^4q f(q,a\mu)}
\label{eq:scale}
\ee
where $f(q,a\mu)$ is the function entering in the loop integral which
defines the constant $C_1(a\mu)$ of eq.~(\ref{eq:zb}):
$C_1(a\mu)=\int d^4q f(q,a\mu)$.%
\footnote{Here we are dealing with multiplicative renormalization, for which
the above recipe is fairly non ambiguous. However, if the operator mixes with
others under renormalization, the recipe is ambiguous, since one can either
evaluate all mixing constants at the same scale $q^\ast$ or obtain each mixing
constant at a separate scale.}
From the above expression we readily see that in general $q^\ast$ depends on
the renormalization scale $\mu$. Here, however, we are only
interested in the finite quantities $Z_V$, $Z_A$ and $Z_S/Z_P$, for which
the $C_i$'s of eq.~(\ref{eq:zunb}) are $\mu$-independent. Consequently,
also $q^\ast$ does not depend on $\mu$.

The connection between the bare coupling $g_0^2$ and the renormalized coupling 
$g_V^2$ defined in eq.~(\ref{eq:vq}) is obtained by calculating the heavy quark 
potential in lattice PT. At lowest order one finds
\be
g_V^2(q) = g_0^2(a) [ 1 - g_0^2(a) 2 \beta_0 \ln(aq/c) + \calo(g^4_0(a)) ]
\label{eq:gs}
\ee
where $\beta_0 = 11/(4\pi)^2$ and $c$ is given by:
\be
c = \pi \exp \Biggl( \frac{4.702}{8 \pi \beta_0} \Biggr)
\ee
There are several possibilities, suggested in \cite{lep_mac}, for
extracting $g_V^2(q^\ast)$ from eq.~(\ref{eq:gs}).
They can be classified as follows:

\underline{1. Purely Perturbative Boosting [PPB]:} This method consists in
using eq.~(\ref{eq:gs}) in order to identify the couplings at scale $q=c/a$:
\be
g^2_V(c/a) = g_0^2(a) + \calo (g_0^4)
\ee 
Then one can perform 2-loop Renormalization Group (RG) running of $g^2_V(c/a)$ 
down to the scale $q^\ast$ in the standard fashion:
\be
\frac{1}{g^2_V(q^\ast)} = \frac{1}{g^2_V(c/a)}
+ 2 \beta_0 \ln (\frac{a q^\ast}{c}) + \frac{\beta_1}{\beta_0}
\ln\biggl (\frac{g_V^2(c/a)}{g^2_V(q^\ast)}\biggr )
\ee
with $\beta_1 = 102/(4\pi)^4$. Thus, having calculated $q^\ast$ (from 
eq.~(\ref{eq:scale})) and $g_V^2(q^\ast)$, we derive $Z_O$ from eq.~(\ref{eq:zb}). 
We call this estimate PPB.

\underline{2. Monte Carlo - Perturbative Boosting [MC-PB($\ln\Box$)]:}
Again this method is based on the relationship between $g_V^2$ and 
$g_0^2$ of eq.~(\ref{eq:gs}).
But now the boosted coupling constant is extracted from the non-perturbative
(Monte Carlo) value of a short distance quantity, such as the average
plaquette \cite{parisi}. The logarithm 
of the plaquette, known as a perturbative series in $g_0^2$, is expressed, with 
the aid of eq.~(\ref{eq:gs}), as a series in $g_V^2$. The leading coefficient of 
this series is an integral of the kind $\int d^4q f_{\Box}(q)$, 
which gives rise, through the criterion of eq.~(\ref{eq:scale}), to the scale 
$q^\ast_\Box$, appropriate for this expansion. From ref.~\cite{lep_mac} 
we have $q^\ast_\Box = 3.41/a$ and the expansion:
\be
\ln(\frac{1}{3} Tr \langle U_\Box \rangle)
= -\frac{1}{3} g_V^2(q^\ast_\Box) [ 1 - 9.46 \cdot 10^{-2}
g_V^2(q^\ast_\Box)
+ \calo(g_V^4) ]
\label{eq:mcpblog}
\ee
The series of the plaquette's logarithm has been preferred to that of the
plaquette itself because, as claimed in ref.~\cite{lep_mac}, its perturbative 
expansion is better behaving. Using the Monte Carlo result for the
average plaquette, eq.~(\ref{eq:mcpblog}) is solved for $g_V^2(q^\ast_\Box)$.
2-loops RG running subsequently yields the desired $g_V^2(q^\ast)$.
With this value for $g_V^2(q^\ast)$, we derive $Z_O$ from eq.~(\ref{eq:zb}).
We call this method MC-PB($\ln\Box$).

\underline{3. Monte Carlo- Perturbative Boosting [MC-PB($\Box$)]:}
Yet another proposal of \cite{lep_mac} consists in combining eqs.~(\ref{eq:gs}), 
at the scale $q=\pi/a$, and the perturbative expansion of the plaquette
\be
\frac{1}{3} Tr \langle U_\Box \rangle =
1 - \frac{1}{3} g_V^2(\pi/a) + 8.71 \cdot 10^{-2} g_V^4(\pi/a)
+ \calo(g_V^6)
\label{eq:p}
\ee
to obtain:
\be
g^2_V(\pi/a)
[ 1 - 0.513 \frac{1}{4\pi} g^2_V(\pi/a) + \calo (g_V^4)]
= \frac{g^2_0(a)}{\frac{1}{3} Tr \langle U_\Box \rangle}
\label{eq:mcpt}
\ee
Once $g^2_V(\pi/a)$ is obtained from the above expression, two loop RG running 
yields $g^2_V(q^\ast)$ and subsequently $Z_O$. We call this estimate
MC-PB($\Box$).

\underline{4. Naive Boosted Perturbation Theory [NBPT]:}
A particularly simple choice of boo\-sted coupling \cite{parisi} is $\tilde g^2 
= g_0^2/
(\frac{1}{3} Tr \langle U_\Box \rangle)$ and the subsequent substitution of 
$g_0$ by $\tilde g^2$ in eq.~(\ref{eq:rlat}). This recipe (a simplification
of eq.~(\ref{eq:mcpt})) is denoted by NBPT. It has been implemented, 
for example, in refs.~\cite{fbs62} and \cite{bknp}. As argued in ref. 
\cite{lep_mac}, Mean Field arguments can also be used in order to support this 
prescription.

Having classified the various boostings, we now comment on them. All these methods 
should be in principle equivalent to lowest order in PT. From the 
field theoretic point of view, PPB is a perfectly legitimate operation, as it 
amounts to a different choice of expansion parameter.
The two MC-PB prescriptions may 
be even superior to PPB in practice, since they possibly incorporate some
non-perturbative and higher order effects. In order to implement these recipes
to the perturbative evaluation of the RC's of interest, we have calculated
the appropriate scale $q^\ast$; this is different for each RC. The values
of $q^\ast$ for $Z_V$, $Z_A$ and $Z_S/Z_P$ are collected in table
\ref{tab:qst}. Some technical details concerning their evaluation are gathered
in the Appendix.

In table \ref{tab:bpt}, we collect the 
values of $Z_O$, obtained from the various boosting recipes listed above.
We see that the variation of the boosted coupling, due to the choice of
boosting recipe, does not exceed $20\%$, and that of the various RC's is about 
$10\%$. These variations reflect the systematic error introduced by the 
truncation of the perturbative series. 
In the next section, we will compare boosted results to the ones obtained non 
perturbatively from WI's. For this comparison we will use the MC-PB$(\ln \Box)$ 
prescription which is considered the optimal choice by the authors of ref. 
\cite{lep_mac}. As will be shown in table \ref{tab:zz}, such a comparison 
reveals a general trend for BPT to 
reduce the magnitude of the $\calo (g_0^4)$ systematic error.
However, in some cases this error remains appreciable (more than $10\%$). 

\begin{table}
\centering
\begin{tabular}{|c|c|c|c|c|c|}\hline
\multicolumn{3}{|c|}{Wilson} & \multicolumn{3}{|c|}{Clover} \\ \hline
$aq^\ast(Z_V)$ & $aq^\ast(Z_A)$ & $aq^\ast(Z_S/Z_P)$ &
$aq^\ast(Z_V)$ & $aq^\ast(Z_A)$ & $aq^\ast(Z_S/Z_P)$ \\ \hline 
  2.4 & 2.6 & 1.9 & 2.7 & 1.2 & 3.2 \\ \hline
\end{tabular}
\caption{\it The values of the scale $aq^\ast$, characteristic of the boosted
perturbative evaluation of $Z_V$, $Z_A$ and $Z_S/Z_P$. The scales are shown
for both the Wilson and Clover actions.}
\label{tab:qst}
\end{table}

\begin{table}
\centering
\begin{tabular}{|c|c c|c c|c c|}\hline
\multicolumn{7}{|c|}{Wilson - $\beta$=6.0} \\ \hline
        & $g^2$ & $Z_V$ & $g^2$ & $Z_A$ & $g^2$ & $Z_S/Z_P$ \\ \hline 
SPT               & 1.00 & 0.83 & 1.00 & 0.87 & 1.00 & 1.10 \\
PPB               & 1.81 & 0.69 & 1.77 & 0.76 & 1.93 & 1.25 \\
MC-PB(ln$\Box$)   & 2.14 & 0.63 & 2.08 & 0.72 & 2.32 & 1.34 \\
MC-PB($\Box$)     & 1.97 & 0.66 & 1.92 & 0.74 & 2.13 & 1.29 \\
NBPT              & 1.68 & 0.71 & 1.68 & 0.77 & 1.68 & 1.20 \\ \hline
\multicolumn{7}{|c|}{Clover - $\beta$=6.0} \\ \hline
        & $g^2$ & $Z_V$ & $g^2$ & $Z_A$ & $g^2$ & $Z_S/Z_P$ \\ \hline 
SPT               & 1.00 & 0.90 & 1.00 & 0.98 & 1.00 & 1.20 \\
PPB               & 1.75 & 0.83 & 2.25 & 0.96 & 1.67 & 1.42 \\
MC-PB(ln$\Box$)   & 2.05 & 0.80 & 2.81 & 0.95 & 1.95 & 1.54 \\
MC-PB($\Box$)     & 1.90 & 0.81 & 2.52 & 0.96 & 1.81 & 1.48 \\
NBPT              & 1.68 & 0.83 & 1.68 & 0.97 & 1.68 & 1.43 \\ \hline
\end{tabular}
\caption{\it Wilson and Clover action RC's obtained with different boosting 
procedures (see text for notation). The RC's are shown at $\beta = 6.0$.}
\label{tab:bpt}
\end{table}

\section{Lattice WI's with Wilson Fermions}
\label{sec:gen}

In this section we review the WI's which follow from the chiral flavour
symmetry of the QCD action. They allow to compute the finite RC's of
lattice operators in a non-perturbative way. Several properties of these RC's 
can be derived from the WI's. Although the results presented here are well
known (see \cite{ks,boc}), we give particular emphasis to the conditions 
(e.g. continuum limit, chiral limit etc.) under which these results are exact. 
This is essential to the understanding of the sources of systematic error in 
practical applications of the WI's, which form the basis of our analysis.
The reader familiar with refs.~\cite{ks,boc}) may skip this section.

We consider local $SU_L(N_f)\times SU_R(N_f)$ chiral transformations
of the fermionic fields:
\bea
&& \delta \psi(x)  =  i \left[ \alpha_V^a (x) \frac{\lambda^a}{2} +
\alpha_A^a (x) \frac{\lambda^a}{2} \gamma_5 \right] \psi (x)
\nonumber \\
&& \delta \bar \psi (x) = -i \bar \psi (x) \left[ \alpha_V^a (x)
\frac{\lambda^a}{2} - \alpha_A^a (x) \frac{\lambda^a}{2} \gamma_5 \right]
\label{eq:vtr}
\eea
and examine the WI's derived separately in the vector and axial cases.

\subsection{Vector WI's on the Lattice}

With degenerate quark masses (i.e. $M_0$ in eq.~(\ref{eq:sf}) proportional to
the unit matrix) global vector transformations are a symmetry of the action. 
From the corresponding local transformations (eqs.~(\ref{eq:vtr}) with
$\alpha_A^a = 0$) the following vector WI can be derived \cite{ks}:
\bea
i \langle \frac{\delta O(x_1,\dots,x_n)}{\delta \alpha_V^a (x)} \rangle & = &
a^4 \sum_\mu \nabla_x^\mu \langle O(x_1,\dots,x_n) \widetilde V^a_\mu(x) \rangle +
\nonumber \\
&& + a^4 \langle O(x_1,\dots,x_n) \bar \psi(x)
[\frac{\lambda^a}{2},M_0] \psi(x) \rangle
\label{eq:vwi}
\eea
where $O(x_1,\dots,x_n)$ is any multi-local operator consisting of quark and
gluon fields at different space-time points ($x_1 \neq x_2 \neq \dots x_n$). 
$a \nabla_x^\mu f(x) = (f(x) - f(x-\mu))$ is an asymmetric lattice derivative, and
\bea
\widetilde V_\mu^a(x) & = &
\frac{1}{2} [\bar \psi (x) ( \gamma_\mu - 1) U_\mu (x) \frac{\lambda^a}{2}
\psi(x + \mu) + \nonumber \\
&& + \bar \psi (x+\mu) ( \gamma_\mu + 1) U_\mu^{\dagger}(x) \frac{\lambda^a}{2}
\psi(x)]
\label{eq:vtilde}
\eea
is a point-split vector current. By keeping the point $x$ separate from 
the points $x_1,\dots, x_n$ and by taking the limit of degenerate bare 
quark masses, we see that eq.~(\ref{eq:vwi}) leads to the conservation of the 
point split lattice vector current, $ \nabla_x^\mu \widetilde V^a_\mu(x) = 0$. 
This also implies that the standard local vector current $V^a_\mu(x) = \bar
\psi(x)\dfrac{\lambda}{2} \gamma_\mu \psi (x)$ is not conserved on the lattice.

For the specific choice $O(x_1,x_2) = \psi (x_1) \bar \psi(x_2)$, the vector
WI (\ref{eq:vwi}) becomes
\bea
\sum_\mu \nabla_x^\mu \langle \psi (x_1) \widetilde V^a_\mu (x)
\bar \psi (x_2) \rangle =
\langle \psi (x_1) \bar \psi (x) [M_0 , \frac{\lambda^a}{2} ] \psi (x)
\bar \psi (x_2) \rangle +
\nonumber \\
+ \delta(x_2 - x) \langle \psi (x_1) \bar \psi (x_2) \rangle
\frac{\lambda^a}{2}
- \delta(x_1 - x) \frac{\lambda^a}{2}
\langle \psi (x_1) \bar \psi (x_2) \rangle
\label{eq:vwiud}
\eea
Let us consider the case of degenerate quark masses ($M_0 \propto 1$)
and place the vector current at the origin ($x = 0$). By Fourier -
transforming the above equation and amputating the corresponding Green's 
function $G_{\widetilde V}^{\mu,a} = \langle \psi (x_1) \widetilde V^{\mu,a}(0) 
\bar \psi (x_2) \rangle$ (c.f. eqs.~(\ref{eq:gos}), (\ref{eq:gp}) and
(\ref{eq:amp})), we obtain
\be
\sum_\mu \left[\frac{1 - \exp(-iaq_\mu)}{a}\right] 
\Lambda_{\widetilde V}^{\mu,a} \biggl ( p+\frac{q}{2},p-\frac{q}{2} \biggr )
= - \cals \biggl (p+\frac{q}{2}\biggr )^{-1} \frac{\lambda^a}{2}
+ \frac{\lambda^a}{2} \cals \biggl (p-\frac{q}{2}\biggr )^{-1}
\label{eq:vwift}
\ee
This is the lattice version of the vector WI. In the continuum limit, it is 
a formal expression relating  bare quantities which are divergent. In order to 
render them meaningful, it is essential to go over to renormalized expressions. 
With the RC's strictly defined in the limit of vanishing cutoff, we 
express the bare quantities of eq.~(\ref{eq:vwift}) in terms of their
renormalized counterparts, given in eqs.~(\ref{eq:cp}) and (\ref{eq:cl}).
Thus, in the limit $a \rightarrow 0$, we obtain:
\be
\sum_\mu iq_\mu Z_{\widetilde V}^{-1} \hat \Lambda_V^{\mu,a} 
\left( p+\frac{q}{2}, p-\frac{q}{2} \right) =
- \hat \cals \biggl (p+\frac{q}{2}\biggr )^{-1} \frac{\lambda^a}{2}
+ \frac{\lambda^a}{2} \hat \cals \biggl (p-\frac{q}{2}\biggr )^{-1}
\label{eq:vwift2}
\ee
However, the definitions of the RC's of eqs.~(\ref{eq:cp}) and (\ref{eq:cl}),
used to derive the above result, are only formal. A concrete definition
implies the choice  of a renormalization scheme. In the case of the vector
current, it is crucial to  impose that the renormalized operator
$\hat V^a_\mu$ satisfy the nominal WI of the 
continuum theory. From this requirement on eq.~(\ref{eq:vwift2}) we obtain that
the RC of the point split conserved vector current (\ref{eq:vtilde}) is 
\cite{ks,boc}:
\be
Z_{\widetilde V}=1
\ee
This result, which is exact in the limit $a \rightarrow 0$, guarantees a proper 
definition of the vector charge and the validity of current algebra. Thus, the 
vector symmetry of the theory survives renormalization.

The local vector current $V_\mu^a(x) = \bar \psi (x) \dfrac{\lambda^a}{2} 
\gamma_\mu \psi(x)$ is not conserved on the lattice. However, it can be shown 
that its ME's differ from those of the conserved current by finite
contributions. The argument goes as follows: first we express the the conserved 
current as
\be
\label{eq:expv}
\widetilde V^a_\mu(x) = V^a_\mu(x) + 
\frac{a}{2}
[\bar \psi (x) (\gamma_\mu - 1) \frac{\lambda^a}{2} \dcr_\mu \psi(x)
+\bar \psi (x) \dcl_\mu (\gamma_\mu + 1) \frac{\lambda^a}{2} \psi(x)]
\ee
where we have used the lattice asymmetric covariant derivatives
\bea
&& a \dcr_\mu \psi (x) = U_\mu(x) \psi(x+\mu) - \psi(x) \nonumber \\
&& a \bar \psi (x) \dcl_\mu = \bar \psi(x+\mu) U_\mu^\dagger(x) -
\bar \psi(x)
\eea
The second term on the r.h.s. of eq.~(\ref{eq:expv}) is a dimension-4 operator 
(we call it $\Delta^a_\mu$) multiplied by the lattice spacing. For definitiveness, 
we now consider the Green's function $\Lambda_{\widetilde V}(p)$ of 
eq.~(\ref{eq:amp}). From eq.~(\ref{eq:expv}) it follows that
\be
\Lambda_{\widetilde V}(p) = \Lambda_V(p) + a \Lambda_\Delta (p)
\label{eq:expg}
\ee
The term $a \Lambda_\Delta (p)$ vanishes at tree-level in the naive continuum 
limit. Beyond tree-level, however, this term contributes, due to power 
divergences induced by the mixing with lower dimensional operators. The mixing 
with operators of equal dimension gives at most logarithmic divergences $\ln(ap)$; 
thus in the continuum limit such contributions vanish like $a \ln(ap)$. As far as 
lower dimensional operators are concerned, $\Delta^a_\mu$ only mixes with the
dimension-3 vector current $V^a_\mu$. This gives rise to power
divergences $a^{-1}$, without logarithms%
\footnote{The absence of logarithms in such mixing has been explicitly
demonstrated in ref.~\cite{curci} for the axial current at all orders
in PT; the situation is analogous for the vector current.}.
Thus, the term $a \Lambda_\Delta$ gives finite contributions, which combine with 
the renormalization of $\Lambda_V$ to give the known non-renormalization exact 
result $Z_{\widetilde V} = 1$ of the l.h.s. of eq.~(\ref{eq:expg}). Consequently, 
the term $\Lambda_V$ on the r.h.s. has a finite renormalization:
\be
Z_V (g_0^2) \ne 1
\ee 
Being finite, $Z_V$ can only depend on the coupling $g_0^2$. Note that 
eq.~(\ref{eq:vwift2}) (and any other vector WI) can now be expressed in terms of 
the non-conserved local current; we simply substitute $\hat \Lambda_V$ by $Z_V 
\Lambda_V$. This is the basis of the non-perturbative WI calculation of $Z_V$, 
detailed in sect. \ref{sec:wi}.

One can also construct useful WI's based on hadronic correlation functions.
For definitiveness, we consider the case in which the operator 
$O(x_1,\dots,x_n)$ of eq.~(\ref{eq:vwi})
is $O(x_1,x_2) = P^{12}(x_1)P^{31}(x_2)$, where $P^{12} = \bar \psi_1 
\gamma_5 \psi_2$. With $x \neq x_1,x_2$, so that contact terms do not contribute, 
eq.~(\ref{eq:vwi}) becomes
\be
\sum_\mu \nabla_\mu^x \langle P^{12}(x_1) \widetilde V_\mu^{23} (x) P^{31} 
(x_2) \rangle = (m_2 - m_3) \langle P^{12}(x_1) S^{23}(x) P^{31} (x_2) \rangle
\label{eq:vwipvp}
\ee
In analogy to eq.~(\ref{eq:vwift2}), we can use eqs.~(\ref{eq:mrzm}) and 
(\ref{eq:zodef}) in order to express the above identity in terms of the
renormalized 
densities, current and masses. By requiring that the renormalized quantities
obey the nominal vector WI, we then find that the product of the quark mass 
times the scalar density is renormalization group invariant. Thus,
we obtain for the RC of the scalar density $S(x)$:
\be
Z_S = Z_m^{-1}
\label{eq:zsm}
\ee

\subsection{Axial WI's on the Lattice}

Far less straightforward is the implementation of the axial symmetry
with Wilson fermions, because of the presence of the chiral symmetry
breaking Wilson term in the action. This topic has been dealt with in
refs.~\cite{ks,boc}. The basic idea is that, by imposing suitable
renormalization conditions, PCAC is recovered in the continuum. The
axial WI, obtained from eq.~(\ref{eq:vtr}) with $\alpha_V^a = 0$, is
\bea
&& i \langle \frac{\delta O(x_1,\dots,x_n)}{\delta \alpha_A^a (x)} \rangle =
a^4 \sum_\mu \nabla_x^\mu \langle O(x_1,\dots,x_n) \widetilde A^a_\mu(x) 
\rangle - \nonumber \\
&& \qquad - a^4 \langle O(x_1,\dots,x_n) \bar \psi(x)
\{ \frac{\lambda^a}{2}, M_0 \} \gamma_5 \psi(x) \rangle
- a^4 \langle O(x_1,\dots,x_n) X^a(x) \rangle
\label{eq:awi}
\eea
where $\widetilde A_\mu^a(x)$ is a bilinear point-split axial current given by
\bea
\widetilde A_\mu^a(x) & = &
\frac{1}{2} [\bar \psi (x) ( \gamma_\mu \gamma_5) U_\mu (x) 
\frac{\lambda^a}{2} \psi(x + \mu) + \nonumber \\
&& + \bar \psi (x+\mu) ( \gamma_\mu \gamma_5) U_\mu^{\dagger}(x)
\frac{\lambda^a}{2} \psi(x)]
\label{eq:atilde}
\eea
The term $X^a$ in the above WI is the variation of the Wilson
term under axial transformations:
\be
X^a(x) = -\frac{1}{2} a 
\left[\bar \psi(x) \frac{\lambda^a}{2} \gamma_5 \dcr^2 \psi(x) +
\bar \psi(x) \dcl^2 \frac{\lambda^a}{2} \gamma_5 \psi(x) \right]
\ee
where
\bea
&& a^2 \dcr^2 \psi (x) = \sum_\mu \left[ U_\mu(x) \psi(x+\mu) +
U_\mu^\dagger(x-\mu) \psi(x-\mu) - 2 \psi(x) \right]
\nonumber \\
&& a^2 \bar \psi (x) \dcl^2 = \sum_\mu \left[ \bar \psi(x+\mu) 
U_\mu^\dagger (x) + \bar \psi(x-\mu) U_\mu (x-\mu) - 2 \bar \psi(x) \right]
\eea
Unlike the vector current case, $X^a$ cannot be cast in the form of a
four-divergence.

$X^a$ is a dimension-4 operator which, in the naive continuum limit
vanishes, being of the form ($a \times$dimension-5 operator). However, it 
has divergent ME's beyond tree-level. Its mixing with operators 
of equal and lower dimensions, worked out in ref.~\cite{boc}, can be expressed 
as follows:
\be
\overline X^a(x) = X^a(x) +
\bar \psi(x) \{ \frac{\lambda^a}{2},\overline M \} \gamma_5 \psi(x)
+ (Z_{\widetilde A} -1) \nabla_x^\mu \widetilde A^a_\mu(x)
\label{eq:xbar}
\ee
where naive dimensional arguments tell us that the mixing constant
$Z_{\widetilde A}(g_0^2,am)$ is finite, whereas $\overline M(g_0^2,M_0)$ 
diverges linearly like $a^{-1}$. Logarithmic divergences have been shown
to be absent at all orders in PT \cite{curci}. The mixing constants 
$Z_{\widetilde A}(g_0^2,am)$ and $\overline M(g_0^2,M_0)$, and therefore
$\overline X^a(x)$, are defined 
by requiring that the renormalized axial current satisfies the nominal 
(continuum) axial WI. This amounts to a renormalization condition which
ensures that the renormalized theory has the proper chiral symmetry.
For example, with $O(x_1,x_2) = \psi (x_1)
\bar \psi (x_2)$, eq.~(\ref{eq:awi}) becomes
\bea
&& \nabla^\mu_x \langle \psi(x_1) Z_{\widetilde A} {\widetilde A}^a_\mu(x) 
\bar \psi (x_2) \rangle = \nonumber \\
&& \qquad = \langle \psi(x_1) \bar \psi (x)
\{ \frac{\lambda^a}{2},(M_0-\overline M) \}
\gamma_5 \psi(x) \bar \psi (x_2) \rangle
+ \langle \psi(x_1) \overline X^a(x) \bar \psi (x_2) \rangle -
\nonumber \\
&& \qquad - \delta (x-x_2) \langle \psi (x_1) \bar \psi (x_2) \gamma_5 
\rangle \frac{\lambda^a}{2}
- \delta (x-x_1) \frac{\lambda^a}{2}
\langle \gamma_5 \psi (x_1) \bar \psi (x_2) \rangle
\label{eq:zaqb}
\eea
where $X^a$ has been replaced by $\overline X^a$. We now take the continuum 
limit of the above equation, with the aim of obtaining from it the nominal 
axial WI for the renormalized quantities. We immediately see that this 
identity can be recovered from eq.~(\ref{eq:zaqb}) provided that the 
following renormalization condition is imposed on $\overline X^a$:
\be
\lim_{a \rightarrow 0} \langle \psi (x_1) \overline X^a(x) \bar \psi (x_2)
\rangle = 0
\label{eq:qqrc}
\ee
Moreover, the chiral limit is to be defined as the value $M_C$ of $M_0$
for which the difference $M_C - \overline M (g_0^2,M_C)$ vanishes.
Thus, in the chiral limit, eq.~(\ref{eq:zaqb}) becomes
\bea
\label{eq:zaqr}
&& \nabla^\mu_x \langle \psi(x_1) Z_{\widetilde A} {\widetilde A}^a_\mu(x) 
\bar \psi (x_2) \rangle = \\
&& \qquad - \delta (x-x_2) \langle \psi (x_1) \bar \psi (x_2) \gamma_5 
\rangle \frac{\lambda^a}{2}
- \delta (x-x_1) 
\frac{\lambda^a}{2} \langle \gamma_5 \psi (x_1) \bar \psi (x_2) \rangle
\nonumber
\eea
Expressing the quark fields in terms of their renormalized counterparts, we 
immediately see that the nominal axial WI is recovered up to vanishing cutoff 
effects, provided that we 
interpret the product $\hat A^a_\mu = Z_{\widetilde A} \widetilde A^a_\mu$ 
as the renormalized axial current. Note that $Z_{\widetilde A}(g_0^2,am)$, 
having been fixed in the continuum limit by eq.~(\ref{eq:qqrc}), does 
not depend on $am$ (only a regular dependence on $am$ permits a well defined 
chiral limit). Thus, it is a finite renormalization of the form 
$Z_{\widetilde A}(g_0^2)$.

The above arguments rest on the assumption that the chiral point $M_C$,
defined after eq.~(\ref{eq:qqrc}), is consistent with other definitions (e.g.
the vanishing of the quark mass in PT or of the pion mass). This
has been verified both in 1-loop PT \cite{boc} and non-perturbatively
(e.g. \cite{MARTIR}).

In practice it turns out to be more convenient to work with the lattice
local axial current $A_\mu^a(x)$. We can show, in a fashion analogous to 
the case of the vector current (c.f. the power counting argument based on 
eqs.~(\ref{eq:expv}) - (\ref{eq:expg})), that $A_\mu^a(x)$ has a finite 
normalization constant $Z_A$. Thus we have $\hat A^a_\mu =
\lim_{a \rightarrow 0} [Z_{\widetilde A} \widetilde A^a_\mu] = 
\lim_{a \rightarrow 0} [Z_A A^a_\mu]$. From now on, 
the combination $Z_{\widetilde A} \widetilde A^a_\mu$ will always be 
substituted by $Z_A A^a_\mu$ wherever it appears in a WI. Also analogous to 
the vector current case is the lack of mass dependence of these constants.
We therefore have:
\begin{equation}
Z_{\widetilde A}(g_0^2) \,\, , \,\,  Z_A(g_0^2) \,\, \neq \,\, 1
\end{equation}

Hadronic axial WI's are also very useful. In eq.~(\ref{eq:awi}) we consider 
the operator $O(x_1) = P^{21}(x_1) = \bar \psi_2(x_1) \gamma_5 \psi_1(x_1)$
and $x \ne x_1$, so that contact terms do not contribute. The resulting WI is
\bea
&& Z_A \sum_\mu \nabla^\mu_x \langle A^{12}_\mu(x) P^{21} (x_1) \rangle = 
\langle \overline X^{12} (x) P^{21} (x_1) \rangle +
\nonumber \\
&& \qquad + \bigl[ m_{01} + m_{02} - \overline m_1 -
\overline m_2 \bigr]
\langle P^{12}(x) P^{21}(x_1) \rangle
\label{eq:vwipap}
\eea
where $\overline m_i(g_0^2,M_0)$ is the $i^{th}$ diagonal element of
$\overline M(g_0^2,M_0)$. Note that we have replaced $X$ by $\overline X$ 
according to eq.~(\ref{eq:xbar}). The condition (\ref{eq:qqrc}) implies that 
correlation function of the operator $\overline X$ vanishes in the continuum
limit, except for localized contact terms. Thus, when the points 
$x_1$ and $x$ are kept separate, $\langle \overline X(x) 
\hat P (x_1) \rangle$ vanishes for $a \rightarrow 0$. We now write in
eq.~(\ref{eq:vwipap}) the renormalized quark mass as:
\be
m_R = \overline Z_m [ m_0 - \overline m(m_0)] =
\overline Z_m [ m_0 - m_C 
- \frac{\partial \overline m}{\partial m_0}
\Bigg \vert _{m_C} (m_0 - m_C) + ...]
\label{eq:mr2}
\ee
where we have used the property $\overline m(m_C) = m_C$.
Upon expressing eq.~(\ref{eq:vwipap}) in terms of 
renormalized quantities and requesting the validity of the axial WI in the 
continuum limit, we obtain for the RC $Z_P$ of the pseudoscalar density: 
\be
Z_P = 1/{\overline Z}_m
\label{eq:zpm}
\ee

Note that from the WI of eq.~(\ref{eq:vwipap}) we can only determine the 
combination
\be
2 \rho^{12} = Z_A^{-1} 
[m_{01}+m_{02} - \overline m_1 - \overline m_2 ] =
\frac {\sum_\mu \nabla_\mu^x \langle A^{12}_\mu(x) P^{21}(x_1) \rangle}
{\langle P^{12}(x) P^{21}(x_1) \rangle}
\label{eq:2rho}
\ee
but not $Z_A$. Again, the above is true up to vanishing cutoff effects.

A very useful WI can be derived taking $O(x_1,x_2) = A^b_\nu(x_1) 
V^c_\rho(x_2)$:
\bea
&& Z_A \nabla^\mu_x \langle A^a_\mu(x) A^b_\nu(x_1) V^c_\rho(x_2) \rangle
\nonumber \\
&& \qquad = \langle \bar \psi(x) \{ \frac{1}{2} \lambda^a,(M_0 - 
\overline M) \} \gamma_5 \psi(x) A^b_\nu(x_1) V^c_\rho(x_2) \rangle
+ \langle \overline X^a(x) A^b_\nu(x_1) V^c_\rho(x_2) \rangle +
\nonumber \\
&& \qquad +i f^{abd} \delta (x-x_1) \langle V^d_\nu(x_1) V^c_\rho(x_2) 
\rangle +i f^{acd} \delta (x-x_2)
\langle A^b_\nu(x_1) A^d_\rho(x_2) \rangle
\label{eq:aav}
\eea
where again use of eq.~(\ref{eq:xbar}) has been made. By symmetry
arguments the contact terms to which $\overline X^a$ gives rise must
have the form
\bea
&& \langle \overline X^a(x) A^b_\nu(x_1) V^c_\rho(x_2) \rangle =
-i c_1 f^{abd} \delta (x-x_1)
\langle V^d_\nu(x_1) V^c_\rho(x_2) \rangle +
\nonumber \\
&& \qquad +i c_2 f^{acd} \delta (x-x_2)
\langle A^b_\nu(x_1) A^d_\rho(x_2) \rangle + \dots
\label{eq:xk1k2}
\eea
where the dots stand for localized Schwinger terms, which will vanish after
integration over $x$ (we always keep $x_1 \neq x_2$). We now proceed as 
follows: (I) rewrite eq.~(\ref{eq:aav}), using eq.~(\ref{eq:xk1k2}) and 
expressing 
all bare quantities in terms of the renormalized ones; (II) recall that the 
RC $\overline Z_m$ of $M_0 - \overline M$ is the inverse of the RC $Z_P$ 
(c.f. eq.~(\ref{eq:zpm})); (III) require, as always, that the renormalized 
quantities obey the nominal axial WI. Thus we obtain
\bea
&& c_1 = 1 -\frac{Z_V}{Z_A} \nonumber \\
&& c_2 = \frac{Z_A}{Z_V} - 1
\label{eq:k1k2}
\eea
The above conditions ensure the recovery of the canonical WI in the
continuum limit. This WI is the basis of the non-perturbative calculation of 
$Z_A$ in the fashion of refs.~\cite{mm,clvwi}, where eqs.~(\ref{eq:2rho}) -
(\ref{eq:k1k2}) are combined into
\bea
&& \langle [\nabla^\mu_x A^a_\mu (x) - \bar \psi (x)
\{ \frac{\lambda^a}{2} , \rho \} \gamma_5 \psi (x) ]
A^b_\nu(x_1) V^c_\rho(x_2) \rangle =
\nonumber \\
&&  +i f^{abd} \delta (x-x_1)
\langle V^d_\nu(x_1) V^c_\rho(x_2) \rangle \frac{Z_V}{Z_A^2}
+i f^{acd} \delta (x-x_2)
\langle A^b_\nu(x_1) A^d_\rho(x_2) \rangle \frac{1}{Z_V}
\label{eq:xav}
\eea
with $\rho = Z_A^{-1} (M_0 - \overline M)$ a matrix generalization of 
eq.~(\ref{eq:2rho}). Performing the integration over $x$, and integrating 
over $\vec x_1$ (recall that $x_1 \neq x_2$ is necessary in order to 
eliminate Schwinger terms) we arrive at:
\bea
&& \int dx \int d \vec x_1
\langle \bar \psi (x) \{ \frac{1}{2}\lambda^a, \rho \} \gamma_5 \psi (x)
A^b_\nu(x_1) V^c_\rho(x_2) \rangle =
\nonumber \\
&& \qquad -i \frac{Z_V}{Z_A^2} f^{abd} \int d \vec x_1
\langle V^d_\nu (x_1) V^c_\rho (x_2) \rangle
-i \frac{1}{Z_V} f^{acd} \int d \vec x_1
\langle A^b_\nu(x_1) A^d_\rho(x_2) \rangle
\label{eq:pav}
\eea
Note that this equation is only valid away from the chiral limit, where the 
integral over $x$ of the total divergence $\nabla^\mu_x A^a_\mu (x)$ 
vanishes. At zero quark mass, the term containing the total divergence
contributes because of the presence of massless Goldstone bosons.

The same arguments may be repeated for the operator $O(x_1,x_2)= S^g(x_1)
P^h(x_2)$. By imposing that the renormalized operators $\hat S^f(x_1)$ and 
$\hat P^f(x_2)$ satisfy the (integrated) nominal Ward identity, we arrive at
\bea
&& \int d^4x\int d^3\vec y \ \langle \bar \psi(x) \{\frac{1}{2} 
\lambda^f,\rho\} \psi(x) S^g(x_1 )P^h(x_2)\rangle
= \nonumber\\
&& \quad \frac{Z_P}{Z_A Z_S} d^{fgl}
\int d^3\vec x_1 \ \langle P^l(x_1) P^h(x_2)\rangle
+\frac{Z_S}{Z_A Z_P} d^{fhl}
\int d^3\vec x_1 \ \langle S^g(x_1) S^l(x_2)\rangle
\label{eq:wisp}
\eea
where the $d$'s are defined in eq.~(\ref{eq:ds}). The WI of 
eq.~(\ref{eq:wisp}) is valid if $f\neq g$ and $f\neq h$, otherwise there are 
additional terms containing flavour-singlet currents on the right hand side.
Analogous arguments to the ones used previously, show that the 
ratio $Z_S/Z_P$, being obtained from a WI, is a finite function of $g_0^2$. 
Thus, the densities $S$ and $P$ have the same anomalous dimension.

Another way of obtaining the ratio $Z_P/Z_S$ is found by combining
eqs.~(\ref{eq:mrzm}), (\ref{eq:zsm}), (\ref{eq:mr2}) and
(\ref{eq:zpm}): 
\be
\frac{Z_P}{Z_S} = \frac{m_0 - \overline m(m_0)}{m_0-m_C} = 1 -
\frac{\partial \overline m (g_0^2,m_0)}{\partial m_0} \Bigg \vert
_{m_0=m_C} + \dots
\label{eq:zszpm}
\ee
where $m_0 - \overline m = 2 \rho Z_A$.

In this section we have reviewed the lattice vector and axial WI's, 
emphasizing that they explicitly determine finite RC's such as $Z_V$, $Z_A$ 
and the ratio $Z_S/Z_P$. We have repeatedly stressed that these results are 
strictly true up to vanishing $\calo(a)$ terms. In the next section we will use 
the above WI's in order to determine non-perturbatively the RC's. A crucial
observation is that these computations are performed at fixed finite cutoff 
$a$. In this case, the continuum limit is never taken, and an apparent 
``dependence" of $Z_O$ on $a$ is observed. This ``dependence" is a 
discretization (systematic) error, which will eventually drop out in the 
continuum limit. Minimizing this error is the focal point of the Clover 
action improvement, discussed in subsection \ref{sec:clover}.

\section{Evaluation of RC's from WI's}
\label{sec:wi}

The most accurate method for a non-perturbative determination of RC's is 
based on chiral WI's. Only finite RC's can be extracted from them, such as 
$Z_V$, $Z_A$ and the ratio $Z_S / Z_P$. We will use the 
general results of the previous section in order to show how they can be 
computed non-perturbatively.

These RC's are determined by requiring that bare correlation functions, 
calculated numerically on the lattice at finite UV cutoff $a^{-1}$,
when renormalized, obey the WI's of the continuum theory.
This requirement, which we have justified in the previous section in the 
$a \rightarrow 0$ limit, should be also valid at very small but finite 
lattice spacing, within a given accuracy. At fixed lattice spacing, 
correlation functions can be computed numerically, providing us with an 
intrinsically non-perturbative estimate of each term of a WI. These WI's can 
then be solved for the several RC's. Clearly, this method is free from the 
errors which affect the perturbative estimates of the RC's (i.e. higher
orders in $g_0^2$). It suffers however, from finite cutoff effects. We shall 
address this question in detail in the rest of this paper.

\subsection{Wilson WI Estimates of Renormalization Constants}
\label{sec:wils} 

First we consider the WI computation of the RC of the vector current, from 
suitable ratios of correlation functions. We shall use the following 
symmetrized form of the conserved current:
\be
V_\mu^{a C} =
\frac{1}{2} [ \widetilde V_\mu^a(x) + \widetilde V_\mu^a (x-\mu) ]
\ee
Being just a symmetrization of $\widetilde 
V_\mu^a$, it has the same RC, $Z_{V^C} = 1$. From the discussion of the 
vector WI's between hadronic correlation functions, we conclude that $Z_V$ 
can be derived from ratios of $n$-point correlation functions; e.g.
\bea
R_\rho (\vec q) & = & \frac{\int d \vec x e^{i\vec q \vec x}
\langle V^C_k (x) M_\rho^{k\dagger}(0) \rangle}
{\int d \vec x e^{i\vec q \vec x}
\langle V_k(x) M_\rho^{k\dagger}(0) \rangle}
 \rightarrow
\frac{\sum_r \epsilon^k_r(\vec q) 
\langle 0 \vert V^C_k (0) \vert \rho_r (\vec q) \rangle}
{\sum_r \epsilon^k_r(\vec q)
\langle 0 \vert V_k (0) \vert \rho_r (\vec q) \rangle} + \dots
\label{eq:rat}
\\
R_0 (\vec q) & = & \frac{\int d \vec x \int d \vec y
e^{i\vec q \vec x}
\langle P_K(0) V^C_0(x) P_D^\dagger(y) \rangle}
{\int d \vec x \int d \vec y e^{i\vec q \vec x}
\langle P_K(0) V_0(x) P_D^\dagger(y) \rangle}
\rightarrow
\frac{\langle K (\vec q) \vert V^C_0 (0) \vert D (\vec 0) \rangle}
{\langle K (\vec q) \vert V_0 (0) \vert D (\vec 0) \rangle} + \dots
\nonumber \\
R_k (\vec q) & = & \frac{\int d \vec x \int d \vec y 
e^{i\vec q \vec x} \langle P_K(0) V^C_k(x) P_D^\dagger(y) \rangle}
{\int d \vec x \int d \vec y e^{i\vec q \vec x}
\langle P_K(0) V_k(x) P_D^\dagger(y) \rangle}
\rightarrow
\frac{\langle K (\vec q) \vert V^C_k (0) \vert D (\vec 0) \rangle}
{\langle K (\vec q) \vert V_k (0) \vert D (\vec 0) \rangle} + \dots
\nonumber \\
R_k^\ast (\vec q) & = & \frac{\int d \vec x \int d \vec y
e^{i\vec q \vec x}
\langle M^\lambda_{K^\ast}(0) V^C_k(x) P_D^\dagger(y) \rangle}
{\int d \vec x \int d \vec y e^{i\vec q \vec x}
\langle M^\lambda_{K^\ast}(0) V_k(x) P_D^\dagger(y) \rangle}
\rightarrow
\frac{\sum_r \epsilon^\lambda_r(\vec q)
\langle K^\ast_r (\vec q) \vert V^C_k (0) \vert D (\vec 0) \rangle}
{\sum_r \epsilon^\lambda_r(\vec q)
\langle K^\ast_r (\vec q) \vert V_k (0) \vert D (\vec 0) \rangle} + \dots
\nonumber
\eea
where $k = 1, 2, 3$ is a spatial index and $\epsilon^\lambda_r(\vec q)$
($\lambda=0, \ldots ,3$) the vector meson polarization vectors. 
For definitiveness, we have specified the quark flavours by considering 
the correlations characteristic of the $\rho$ decay (2-point functions)
and those of the $D \rightarrow K$ and $D \rightarrow K^\ast$ decays
(3-point functions). Thus, the operators are defined as:
$P_K = \bar u \gamma^5 s$; $P_D = \bar u \gamma^5 c$; 
$M_\rho^k = \bar u \gamma^k d$; $M^\lambda_{K^\ast} = \bar u \gamma^\lambda s$. 
The vector currents have a $\bar u - d$ flavour structure in $R_\rho$ and
a $\bar u - c$ flavour structure in all other ratios.
In the above equation, we also
show the asymptotic behaviour of these ratios, obtained at large time
separations, up to contributions from higher excited states.
All these ratios give estimates of $Z_V$ up to terms of $\calo (a)$.

The above ratios represent the most straightforward way to obtain $Z_V$, but 
it is by no means the only one. As an example, we present a method, which 
follows more closely the vector WI's but is analogous to the determination 
of $Z_V$ from the ratio $R_0$. It consists in integrating over all 3-space 
$\vec x$ both sides of eq.~(\ref{eq:vwipvp}). Upon integrating, the spatial
derivative $\nabla^k \widetilde V_k$ vanishes (either because fields die-off 
at infinity or due to periodic boundary conditions of a finite lattice). 
Then we write the WI in terms of $V_0^C$. The result is
\be
\int d \vec x 
\langle P^{12} (x_1) \nabla^0_x V_0^C (x) P^{31} (x_2) \rangle =
\nonumber \\
\frac{1}{2} (m_2 - m_3) \int d {\vec x} 
\langle P^{12} (x_1) [S^{23}(x) + S^{23}(x - \hat 0)] P^{31} (x_2) \rangle
\label{eq:wwii}
\ee
where, upon passing from $\widetilde V_0$ to $V_0^C$, we end up with the sum
of two adjacent scalar densities on the r.h.s. Finally, we substitute $Z_V 
V_0$ for $V_0^C$ in the above and solve for $Z_V$. The implementation of 
eqs.~(\ref{eq:rat}) for the computation of $Z_V$ will be referred to as the 
``Ratio" determination. When WI's are directly used  instead, we will call 
it the WI determination. Both are non-perturbative methods, equivalent in 
the $a \rightarrow 0$ limit.

For the WI determination of $Z_A$ we start from eq.~(\ref{eq:pav}). For
definitiveness, we will consider 3 quark flavours, which may or may not
be degenerate in mass. It is necessary to keep track of the flavour
content of each operator, e.g. $O^{12}_\Gamma(x) = \bar \psi_1(x) \Gamma 
\psi_2(x)$. The RC's do not depend on flavour. We shall however
leave flavour indices on them in order to remind us the origin of
possible discretization errors in their determination (e.g. $\calo
(am_1, am_2)$ for $Z^{12}$). With this flavour content, setting $x_2=0$, 
eq.~(\ref{eq:pav}) becomes
\bea
&& 2 \rho^{12} \int dx \int d \vec x_1
\langle P^{12}(x) A^{31}_\nu (x_1) V^{23}_\rho (0) \rangle =
\nonumber \\
&& \qquad = \frac{Z_V^{23}}{Z_A^{13} Z_A^{12}} \int d \vec x_1
\langle V^{32}_\nu (x_1) V^{23}_\rho (0) \rangle
- \frac{Z_A^{13}}{Z_A^{12} Z_V^{23}} \int d \vec x_1
\langle A^{31}_\nu (x_1) A^{13}_\rho (0) \rangle
\label{eq:auds}
\eea
This WI can be useful in several ways. To simplify matters, in practice we 
will consider the case of flavours 2 and 3 degenerate in mass. This means 
that the pre-factors of the two integrals of the r.h.s. of eq.~(\ref{eq:auds})
simplify to $Z_V^{22}/(Z_A^{12})^2$ and $1/Z_V^{22}$. Thus, by solving the 
above WI at different times $t_1$, we obtain estimates of $Z_V^{22}$ and 
$Z_A^{12}$. Alternatively, $Z_V$ can be calculated from, say, one of the 
ratios of eq.~(\ref{eq:rat}) and used in eq.~(\ref{eq:auds}), from which $Z_A$ 
can be computed. Moreover, with $\nu = \rho = 0$, the integrated correlation
$\int d \vec x_1 \langle V_0(x_1) V_0(0) \rangle$ vanishes, since it is 
proportional to the charge of the vacuum. This means that the remaining WI
\be
2 \rho^{12} \int dx \int d \vec x_1
\langle P^{12}(x) A^{21}_0 (x_1) V^{22}_0 (0) \rangle =
-  \frac{1}{Z_V^{22}} \int d \vec x_1
\langle A^{21}_0 (x_1) A^{12}_0 (0) \rangle
\label{eq:vuds}
\ee
can be used for yet another determination of $Z_V$. Besides $Z_A$ and
$Z_V$, the finite ratio $Z_S/Z_P$ can also be obtained from the
WI of eq.~(\ref{eq:wisp}), by a method analogous to the one used for $Z_A$;
eq.~(\ref{eq:auds}).

So far only WI's of hadronic correlation functions have been examined.
Quark correlation functions such as $\langle \psi(x) V_\mu(0) \bar \psi(y) 
\rangle$ and $\langle \psi(x) A_\mu(0) \bar \psi(y) \rangle$ can also be
used in principle with WI's such as eqs.~(\ref{eq:vwiud}) and
(\ref{eq:zaqb}). However, in numerical simulations the ratios of hadronic 
correlation functions turn out to be more stable than those of quark 
correlation functions. For the latter case, we must work in a fixed gauge 
and the Gribov ambiguity causes increased statistical fluctuations. 
Thus, the determination of $Z_V$ and $Z_A$ 
from quark state correlation functions is expected to be noisier. This has 
been explicitly verified for $Z_A$ in refs.~\cite{clvwi} and \cite{gri}.

\subsection{Clover-Improved WI Estimates of Renormalization Constants}
\label{sec:clover} 

As already pointed out, the WI determination of the finite RC's has the 
advantage of being non-perturbative. In principle, all WI's should give the 
same finite $Z_O$ of a given operator $O$, since the RC's are
characteristic of the operator they renormalize, but are independent of 
the ME's from which they are obtained. In practice, however, 
problems arise from the fact that the RC's, properly defined at vanishing 
cutoff, when calculated from WI's (or Ratios), are obtained at finite cutoff.
At finite lattice spacing, operators mix with higher dimensional (irrelevant) 
operators. This mixing spoils the renormalization properties of 
the lattice operators, which is only recovered in the limit $a \rightarrow 
0$. The contamination from these higher dimensional operators to the 
numerical calculation of RC's depends on the correlation functions (or ME's) 
used. Consequently, the result for $Z_O$ obtained at finite cutoff will 
depend on the correlation functions from which it has been obtained and the
quark mass at which the simulation was performed. This unwanted dependence
signals the presence of systematic errors which are effectively $\calo (a)$ 
in the scaling limit. More specifically, they are $\calo(a\Lambda_{QCD}), 
\calo(am),\calo(aq)$ and higher, where $m$ and $q$ stand for the masses and 
spatial momenta that characterize the process under study. 

In order to reduce this systematic error, several proposals have been
put forward \cite{impr}, \cite{sw}, \cite{heatlie}, \cite{lusch1}.
All are based on the Symanzik improvement programme \cite{sym},
which consists in adding suitable irrelevant dimension-5 operators to the
lattice action. Moreover, $d$-dimensional operators must also be redefined
by the addition of $(d+1)$-dimensional improving operators. All proposals
cited above achieve the elimination of the $\calo(a)$ discretization error
(although at different orders in PT), and are referred to as $\calo(a)$
improvement. The most widely used improved action is the SW/Clover
one \cite{sw}, defined as follows:
\be
S_c = S_f - c_{SW}(g_0^2) a^4 \sum_{x,\mu\nu} \frac{a}{4} \bar \psi(x)
\sigma_{\mu \nu} F_{\mu \nu}(x) \psi(x)
\label{eq:sclov}
\ee
with $F_{\mu \nu}(x)$ the clover-leaf discretization of the field tensor. 
The improving coefficient $c_{SW}(g_0^2)$ can be calculated in PT \cite{naik};
more recently a non-perturbative determination has also been performed
\cite{lusch1}. In this work we limit ourselves to its tree-level value 
$c_{SW} = 1$. At this order, all $\calo (a g_0^{2n} \ln^n a)$ terms, which
are effectively of $\calo (a)$ in the scaling limit ($g_0^2 \sim 1/\ln a$), 
are eliminated from correlation functions. At leading-log level, the 
improvement of local operators can be expressed as a rotation of the 
fermion fields \cite{heatlie}. According to this prescription, improved 
bilinear operators are given by
\be
O^I_\Gamma (x) = \bar \psi^R(x) \Gamma \psi^R(x)
\label{eq:oi}
\ee
where the rotated fields are defined through
\bea 
&& \psi^R(x) = \left[ 1 - \frac{a}{4} \left(\dsr - m_0\right) \right] \psi(x)
\nonumber \\
&& \bar \psi^R(x) = \bar \psi(x) \left[1 + \frac{a}{4}\left(\dsl + m_0\right) 
\right]
\label{eq:psir}
\eea
and the symmetric lattice covariant derivatives are:
\bea
&& a \dr_\mu \psi(x) = \frac{1}{2} [U_\mu(x) \psi(x+\hat \mu) -
U^\dagger_\mu(x-\hat \mu) \psi(x-\hat \mu)]
\nonumber \\
&& a \bar \psi(x) \dl_\mu = \frac{1}{2} [\bar \psi(x+\hat \mu)
U^\dagger_\mu(x) - \bar \psi(x-\hat \mu) U_\mu(x-\hat \mu)]
\eea
The improved operators $O^I$ differ from the original ones by terms proportional 
to the cutoff. Consequently, they have different RC's, say $Z_{O^I}$.

For the Clover action, the quark propagator
$\cals^I (x-y) = \langle \psi(x) \bar \psi(y) \rangle$ satisfies
the equations
\bea
&& \left[ \dsr(x) + m_0 + a \wcr \right] \cals^I (x-y) = \delta (x-y)
\nonumber \\
&& \cals^I (x-y) \left[ - \dsl(y) + m_0 + a \wcl \right] = \delta (x-y)
\label{eq:eqmot}
\eea
The terms $\wcr$ and $\wcl$ of eq.~(\ref{eq:eqmot}) are shorthand notation
for the operators arising from the variation with respect to $\psi$ and
$\bar \psi$ of the Wilson term of the fermionic action, including
the Clover term given by eq.~(\ref{eq:sclov}). The tree-level $\calo(a)$ 
improved quark propagator is $\langle \psi^R(x) \bar \psi^R(y) \rangle$. 
Using eqs.~(\ref{eq:psir}) and (\ref{eq:eqmot}), we can write 
it in terms of $\cals^I(x-y)$:
\be
\langle \psi^R(x) \bar \psi^R(y) \rangle =
\left[ 1 - \frac{a}{2} \dsr (x) \right] \cals ^I (x-y)
\left[ 1 + \frac{a}{2} \dsl (y) \right] + \frac{a}{2} \delta (x-y)
+ \calo(a^2)
\label{eq:iprop}
\ee
where $\calo(a^2)$ terms proportional to $\wcr$ and $\wcl$  have been dropped, 
as they do not affect $\calo(a)$ improvement. In computing correlation functions 
between external on-shell hadron states, the $\delta$-function never contributes. 
Therefore these correlation functions can be expressed directly in terms of the 
effective rotated propagator
\be
\cals^{eff}(x-y) = \left[1-\frac{a}{2} \dsr(x)\right] \cals^I (x-y)
\left[ 1+\frac{a}{2} \dsl(y)\right]
\label{eq:srotz1}
\ee
This is the Clover propagator we will mostly use in our computations. The reason
it contains an $\calo (a^2)$ term in its definition is that it is readily
computable in numerical simulations \cite{MARTIR}. Thus, it has been extensively 
used in several Clover improved lattice QCD computations of on-shell ME's. 
However, upon studying the lattice WI's we find cases for which the space-time 
points of correlation functions are allowed to coincide. The on-shell ME's 
argument is then invalidated and the complete propagator of eq.~(\ref{eq:iprop}) must 
be used, with the $\delta$-function giving rise to contact terms.

More generally, we can add to the improved propagator
$\langle \psi^R(x) \bar \psi^R(y) \rangle$ the two equations of motion
(\ref{eq:eqmot}), multiplied by an arbitrary factor $z$. Up to
$\calo(a^2)$, we then find
\bea
&& \langle \psi^R(x) \bar \psi^R(y) \rangle_z = \nonumber \\
&& \qquad \left[ 1 - \frac{a}{2} \left(z \dsr (x) - (1-z) m_0\right)\right] 
\cals ^I (x-y) \left[ 1 + \frac{a}{2} \left(\dsl (y) z + (1-z) m_0\right)\right]
\nonumber \\
&& \qquad +\frac{a}{2} (2z-1) \delta (x-y) + \calo(a^2)
\label{eq:ipropz}
\eea
In this case, improved on-shell correlation functions can be expressed in terms 
of traces of the effective propagator
\be
\cals^{eff}_z(x-y) = 
\left[ 1 - \frac{a}{2} \left(z \dsr (x) - (1-z) m_0\right)\right] \cals ^I (x-y)
\left[ 1 + \frac{a}{2} \left(\dsl (y) z + (1-z) m_0\right)\right]
\label{eq:srotzz}
\ee
with $z$ a free parameter, which we have at our disposal for optimization 
purposes. Clearly, for different choices of $z$, these ME's (and the 
corresponding improved operators) differ by $\calo(a^2)$ terms. This implies 
that their RC's also vary with $z$. In this work, unless otherwise stated, 
we shall use operators $O^I_\Gamma$ which correspond to $z=1$. 

Although on-shell correlation functions of local improved operators $O^I_
\Gamma$ are now improved up to $\calo(g_0^2a)$, the conserved (point-split)
current is not improved. An improved conserved current is given by 
\cite{masavla}
\be
V_\mu^{CI} (x) = V_\mu^C + \frac{1}{2} a \overline \nabla _\rho
( \bar \psi \sigma_{\rho\mu} \psi )
\label{eq:vci}
\ee
where the extra ``$\sigma_{\rho \mu}$ term" has a symmetric lattice
derivative ($2 a \overline \nabla_\mu f(x) = f(x+\mu) - f(x-\mu)$).

Having improved both local and conserved operators, we can now
use them in the ratios of eq.~(\ref{eq:rat}), in order to obtain estimates
of $Z_{V^I}$ which only suffer from $\calo(a g_0^2)$ discretization errors.
On the other hand, the l.h.s of the WI of eq.~(\ref{eq:wwii}), even when 
expressed in terms of $P^I$ and $V_0^{CI}$, still suffers from
$\calo(a)$ corrections. This is due to the presence of the asymmetric lattice
derivative $\nabla^\mu_x = \partial^\mu_x + \calo(a)$. In order to
improve it, we must express it in terms of the symmetric derivative
$\overline \nabla^\mu_x = \partial^\mu_x + \calo(a^2)$ and make use of
the fact that the conserved currents $V_\mu^C$ and $V_\mu^{CI}$ have the
same divergence ($\overline \nabla^\mu V^C_\mu = \overline \nabla^\mu
V^{CI}_\mu$). Thus we arrive at the following $\calo(a)$ 
improved version of eq.~(\ref{eq:wwii}):
\bea
\label{eq:wwiimp}
&& \int d \vec x
\langle P^{12} (x_1) \overline \nabla^0_x V_0^{CI} (x) P^{31}(x_2) \rangle = \\
&& \qquad = \frac{(m_2 - m_3)}{4} \int d {\vec x} 
\langle P^{12} (x_1)
[2 S^{23}(x) + S^{23}(x + \hat 0) + S^{23}(x - \hat 0)]
P^{31}(x_2) \rangle \nonumber
\eea
In the above equation we have dropped the $I$ superscript from the density
$P$. The factor $1/4$ on the r.h.s. comes from $\overline \nabla^\mu_x$.
We can now express $V_0^{CI}$ as $Z_{V^I} V^I_0$ and solve for
$Z_{V^I}$. This estimate of the RC is free of leading $\calo(a)$ errors. 

We now turn to the WI of eq.~(\ref{eq:auds}) which, expressed
in terms of improved operators, can be used for the determination of the
tree-level $\calo(a)$ improved RC $Z_{A^I}$ of the axial
current. The subtlety is that the integration $\int d^4 x$ of the WI
requires inclusion of the contact terms arising from the $\delta$-function
of the rotated propagator of eq.~(\ref{eq:ipropz}). Thus, for any value of
the $z$-parameter, the Clover improved version of eq.~(\ref{eq:auds}) becomes
\bea
&& 2 \rho^{12} \int dx \int d \vec y
\langle P^{12}(x) A^{31}_\nu (y) V^{23}_\rho (0) \rangle =
(\frac{Z_{V^I}^{23}}{Z_{A^I}^{31} Z_{A^I}^{12}}
- \rho^{12} a \bar z)
\int d \vec y \langle V^{32}_\nu (y) V^{23}_\rho (0) \rangle
\nonumber \\
&& \qquad - (\frac{Z_{A^I}^{13}}{Z_{A^I}^{12} Z_{V^I}^{23}}
- \rho^{13} a \bar z)
\int d \vec y \langle A^{31}_\nu (y) A^{13}_\rho (0) \rangle
\label{eq:audsi}
\eea
where $\bar z = (2z-1)$.
All operators in the above equation, as well as the $2\rho$ factors,
are to be understood as improved quantities. For $2\rho$
this is true provided that it is computed from 
eq.~(\ref{eq:2rho}) with improved operators $A^I_\mu$ and $P^I$ and a 
symmetrized lattice temporal derivative $\overline \nabla^0$.

\subsection{Comparison of the Perturbative and WI Estimates of the 
Renormalization Constants}

From the discussion so far, it should be clear that it is essential to the 
reliability of any WI computation, performed at finite cutoff, that any 
scale with mass dimensions (such as $\Lambda_{QCD}$, masses $m$ and momenta 
magnitudes $q$) must satisfy the condition
\be
\Lambda_{QCD}, m, q \ll a^{-1}
\label{eq:win}
\ee
In an ideal situation ($g_0^2$ and $a$ extremely small) the above condition 
is satisfied and the estimates of the Z's obtained from different correlation 
functions with the WI/Ratio method should all agree. They should also agree
with the Z estimates obtained in PT, carried out at sufficiently high order.
The argument can now be turned upon its head: if the 
$Z_O$'s obtained from WI's or ratios of different correlation functions are 
not in good agreement, this is to be interpreted as a signal that the 
calculation suffers from large discretization errors. In fact, the 
``dependence" of $Z_O$ from the external states and/or the quark mass is a 
criterion used to estimate the finite lattice spacing effects which affect 
the lattice calculations. On the other hand, if the WI/Ratio result for 
$Z_O$ turns out to be reliable, any discrepancies with its PT estimate 
is to be attributed to the truncation of the perturbative series.

To illustrate the point, we collect some results in table \ref{tab:zz},
obtained at a typical value of the lattice coupling constant, $\beta=6.0$. 
\begin{table}
\centering
\begin{tabular}{|c|l|l|l|}\hline
       &     Method       &        Wilson         & Clover \\ \hline \hline
 $Z_V$ & SPT              & 0.83                  & 0.90              \\
       & MC-PB($\ln\Box$) & 0.63                  & 0.80              \\
       & Ratio ($R_\rho$) & 0.57(1) \cite{sach}   & 0.82    [This work] \\
       & Ratio ($R_0$)    & 0.73(3) \cite{sach}   & 0.82    [This work] \\
       & WI (Axial)       & 0.79(4) \cite{mm}     & 0.80(2) [This work] \\
       & NP               & 0.75(5) \cite{schier} & 0.84(1) \cite{np} \\ \hline       
 $Z_A$ & SPT              & 0.87                  & 0.98              \\
       & MC-PB($\ln\Box$) & 0.72                  & 0.95              \\
       & WI (Axial)       & 0.85(7) \cite{mm}     & 1.11(2) [This work] \\
       & NP               & 0.80(5) \cite{schier} & 1.06(8) \cite{np} \\ \hline   
$Z_S/Z_P$ & SPT              & 1.10      & 1.20               \\
          & MC-PB($\ln\Box$) & 1.34      & 1.54                \\
          & WI (Axial)       &  N/A      & 1.64(5) [This work]  \\
          & NP               &  N/A      & 1.6(1)  \cite{np}  \\ \hline
\end{tabular}
\caption{\it
Estimates of the RC's of the vector current ($Z_V$), 
the axial current ($Z_A$) and the ratio of the scalar and pseudoscalar 
densities ($Z_S/Z_P$) from various perturbative and non-perturbative methods 
(see text for labels) at $\beta = 6.0$. All results are obtained at a small
value of the light quark masses ($am \simeq 0.07$). The corresponding 
references are shown in square brackets.}
\label{tab:zz}
\end{table}
For the Wilson case, the presence of large $\calo (a)$ effects is highlighted
by the $20\% - 30\%$ variation in the various non-perturbative estimates of 
$Z_V$. This wide spread of values renders problematic their comparison to the
perturbative estimates (both SPT and BPT). For $Z_A$, the non-perturbative
values are compatible within statistical errors. However, these
errors are too large to allow any conclusion on the accuracy of the two
perturbative determinations.

For Clover fermions, the spread of the different WI estimates is reduced to 
less than $5\%$. This is due to the elimination of $\calo(a)$ terms guaranteed 
by the Clover improvement. This result implies that lattice QCD calculations, 
performed with the Clover action at light quark mass, are significantly 
less affected by the finiteness of the lattice spacing than their Wilson 
counterparts. Note that BPT predictions for $Z_V$ agree well with the
non-perturbative results. However, for the ratio
$Z_S/Z_P$ BPT is less successful.
Moreover, the PT estimate of $Z_A$ is less than 1, whereas the NP value 
exceeds unity. BPT cannot rectify this discrepancy. Thus, $O(g^4)$ terms 
are probably important in these cases. 

So far we have presented results at $\beta = 6.0$. The reason for this
choice is that this coupling is typically used by present-day lattice
QCD groups for calibration purposes. Most studies of an explorative
nature (e.g. RC's from WI's, or from the NP method, with or without Clover 
improvement) have first been carried out at $\beta = 6.0$. Thus a comparison 
of the RC's obtained from all different methods is possible at this value of the
coupling. In subsection \ref{subs:zvw} we also present results for $Z_V$ obtained
from the ratio method with the Wilson action at $\beta=6.4$. Although the
cutoff $a^{-1}$ is considerably increased, we will see that finite cutoff
effects remain sizeable. Thus, our qualitative conclusions on systematic
errors characterizing the Wilson action remain valid at $\beta =6.4$.
Finally, finite RC's with the Clover action (at light quark mass) have also been 
computed at $\beta=6.2$ \cite{ukzs}.

\section{Lattice - Continuum Normalization of Fields and Currents}
\label{sec:KLM}

We now examine a proposal due to Lepage, Mackenzie and Kronfeld \cite{lep}
-\cite{paul} for the removal of $\calo(am)$ systematic errors which are
relevant in lattice calculations with the Wilson action at large quark masses. 
After a critical review of their recipe, we extend it to the Clover 
case. For both the Wilson and Clover cases, we apply it to the WI computation 
of $Z_V$ and $Z_A$ and discuss its merits and shortcomings.

First we set up our notation. So far we have worked with the lattice action
$S = S_f + S_g$ of eqs.~(\ref{eq:sg}) and (\ref{eq:sf}), which depends on 
lattice fields and bare parameters; i.e. it has the form $S [\bar \psi, \psi, 
U_\mu, m_0, g_0^2, a]$. In numerical simulations, it is convenient to work 
with dimensionless fermion fields and the hopping parameter instead of the 
quark mass; i.e. the action used in Monte Carlo simulations 
\bea
S^{LAT} &=& -  \sum _{x,\mu} K 
 \bigl[ \bar \chi (x) (1 - \gamma_\mu) U_\mu (x) \chi (x+\mu)
 + \bar \chi (x+\mu) (1 + \gamma_\mu) U^\dagger_\mu (x) \chi (x) \bigr]
\nonumber \\
&& + \sum_x \bar \chi (x) \chi (x) + S_g
\label{eq:sflat}
\eea
has the form $S^{LAT}[\bar \chi, \chi, U_\mu, K, g_0^2]$, where
$K$ and $\chi$ are defined through
\be
\chi = \sqrt{\frac{a^3}{2K}} \, \psi \qquad , \qquad
\frac{1}{2K} - 4 = a m_0
\label{bare}
\ee
Note that the lattice spacing $a$ has been scaled away in $S^{LAT}$.
All the equations appearing in the previous sections can now be
trivially rewritten in terms of $\bar \chi$ and $\chi$ fields and
masses in lattice units $m_0a$ (or $K$); the dependence on $a$ drops out 
and the equations express relations between dimensionless quantities.
From now on, all lattice quantities (such as quark propagators,
operators and their correlation functions) are to be understood as
defined in terms of the $\bar \chi$ and $\chi$ fields; i.e. they
are (dimensionless) quantities expressed in lattice units.

In order to connect lattice fields to continuum (bare) ones, we must expand
the lattice action for $a \rightarrow 0$ at fixed $g_0$ and $K$. This action
then reduces to the continuum QCD action, $S^{QCD}$, provided we relate the 
continuum field $A_\mu$ to its lattice counterpart $U_\mu$ in the following 
way:
\be
U_\mu = \exp(ia g_0 A_\mu)
\label{bare2}
\ee
The standard prescription for the transcription of any continuum quantity 
on the lattice follows from eqs.~(\ref{bare}) and (\ref{bare2}).
For example, in the naive continuum limit, the dimensionless
lattice quark propagator $\cals(x-y) = \langle \chi(x) \bar \chi(x) \rangle$
becomes its QCD counterpart $\cals^{QCD}(x-y) = \langle \psi(x) \bar 
\psi(x) \rangle$ when normalized as follows:
\be
a^{-3} \,\, 2K \,\, \cals (x-y) \rightarrow \cals^{QCD}(x-y) 
\label{eq:ncls}
\ee
Clearly, such transcriptions are not unique. Modifications of the lattice
quantity by terms which are $\calo (a)$ give equally acceptable transcriptions.
For example, both the lattice local vector current $V_\mu(x) = \bar \chi_1(x) 
\gamma_\mu \chi_2(x)$ and the 
corresponding point-split conserved current $V_\mu^C(x)$ are good transcriptions 
of the vector current obtained from $S^{QCD}$:
\bea
&& a^{-3} \sqrt{2K_1 2K_2} \, V_\mu(x) \rightarrow V_\mu^{QCD}(x) \nonumber \\
&& a^{-3} \sqrt{2K_1 2K_2} \, V_\mu^C(x) \rightarrow V_\mu^{QCD}(x) 
\eea
Note that these relations are just transcriptions of bare QCD quantities
in terms of quantities derived from $S^{LAT}$. They are not relationships 
between lattice and renormalized quantities, which must involve the presence 
of RC's; e.g. $\hat V_\mu = Z_V V_\mu$. In our notation,
the QCD superscript and the ``hat" distinguish the two cases.

\subsection{KLM Factors for the Wilson Action}

The proposal by Lepage, Mackenzie and Kronfeld (KLM) \cite{lep}
-\cite{paul} is a refinement of the above matching. Their aim is to account 
for large $\calo(am)$ corrections in the limit of small $aq$ ($q$ stands 
for the magnitude of any relevant spatial momentum). In practice their recipe 
consists in calculating the matching factor between lattice correlation 
functions and their continuum counterpart in the free theory, for finite 
lattice spacing and vanishing spatial momentum. The claim is that this free
theory correction should account for most of the $\calo(am)$ systematic error 
also in the interacting case. For example, in the free theory, the matching 
between the continuum and lattice quark propagators in ($t,\vec p$)-space 
at $\vec p = \vec 0$ is
\be
\cals^{QCD}(t,\vecp = \vec 0) = a^{-3} 2K (1 + am_0)
\cals (t,\vecp = \vec 0)
\label{eq:klms}
\ee
This implies for the lattice fermion field the matching relation
\be
\psi = \, F_{\psi}(am_0)^{1/2} \,\, \sqrt{\frac{2K}{a^3}} \,\,\, \chi
\ee
where the KLM wave function normalization factor is
\be
F_{\psi} (am_0) = 1+am_0
\ee
This factor tends to 1 in the limit $am_0 \rightarrow 0$ and the standard
matching of eq.~(\ref{bare}) is recovered. We stress that this is a free theory
matching which is due to the granularity of the space-time. It is not a 
renormalization.

A similar procedure applies to all correlation functions. For example
consider the correlation function of the lattice local current between
external quark states at zero spatial momentum,
\be
L_\mu (t_x,t_y) = \sum_{\vec x ,\vec y} \langle
\chi _1(y) V_\mu(x) \bar \chi _2(0) \rangle 
\label{eq:l}
\ee
It can easily be seen that, in the free theory and at finite cutoff, it is 
related to the continuum correlation function by
\be
L_\mu^{QCD}(t_x,t_y) =
a^{-6} 2K_1 2K_2 F_{\psi}(am_{01}) F_{\psi}(am_{02}) L_\mu(t_x,t_y)
\ee
where $am_{01}$ and $am_{02}$ are the bare quark masses of the two distinct 
flavours of the quark fields. Thus, there is no extra KLM normalization, 
apart from the one related to the quark fields. The KLM factor for
the local vector current turns out to be 1. This is also the case of the
local axial current and of any other local bilinear operator. The KLM
factor for the spatial component of the conserved current $V_k^C(x)$ is
also 1. On the other hand, as pointed out in \cite{bern}, the temporal 
component of the conserved current $V_0^C(x)$ requires an extra KLM factor
\be
F_{V_0^C} (am_{01},am_{02}) = \left(1 + \frac{am_{01} + am_{02}}{2}
\right) ^{-1}
\label{eq:fv0c}
\ee
It arises from the matching of
\be
C_0(t_x,t_y) = \sum_{\vec x ,\vec y} \langle
\chi _1(y) V_0^C(x) \bar \chi _2(0) \rangle 
\ee
to $C_0^{QCD}(t_y,t_x)$, which gives:
\be
C_0^{QCD}(t_y,t_x) = a^{-6} 2K_1 2K_2 F_{\psi}(am_{01}) F_{\psi}(am_{02})
F_{V_0^C} (am_{01},am_{02}) C_0(t_y,t_x)
\label{eq:vcbfact}
\ee

It is now trivial to derive the KLM matching factors of hadronic correlation
functions. We just have to consider the number of quark fields and the 
composite operators entering in the correlation function. For example, for the 
2-point function of the vector currents, with the $\rho$-meson quantum numbers, 
we have
\be
\sum_{\vec x} \langle V_k^C(x) V_k^\dagger(0) \rangle^{QCD} = a^{-6}
2K_u 2K_d F_{\psi}(am_{0u}) F_{\psi}(am_{0d})
\sum_{\vec x} \langle V_k^C(x)  V_k^\dagger(0) \rangle
\ee
whereas for the vector 3-point function of the $D \rightarrow K$ decay we have
\bea
&& \sum_{\vec x, \vec y}
\langle P_K(0) V^C_0(x) P^\dagger_D(y) \rangle^{QCD} = \\
&& \qquad = a^{-6}
2K_u 2K_s 2K_c F_{\psi}(am_{0u}) F_{\psi}(am_{0s}) F_{\psi}(am_{0c})
F_{V_0^C} (am_{0s},am_{0c}) \cdot \nonumber \\
&& \qquad \cdot \sum_{\vec x ,\vec y} \langle P_K(0) V^C_0(x) 
P^\dagger_D(y) \rangle
\nonumber
\eea

The KLM factors obtained above depend on the bare quark mass of 
eq.~(\ref{bare}). A possible improvement, which could take into account 
interaction effects, consists in using a more physical definition of the 
quark mass in the KLM factors. A natural choice is, for instance the 
subtracted quark mass $m$ which, in terms of the hopping parameter $K$, is 
given by
\be
am = \frac{1}{2K} - \frac{1}{2K_C} 
\label{eq:2kc}
\ee
with the critical value $K_C$ determined non-perturbatively.
Unless otherwise stated, all KLM factors used in our numerical analysis are 
calculated at the above value of the quark mass (e.g. $F_\psi(am)$,
$F_{V_0^C}(am)$ etc.).
Another proposal has been made in \cite{lep_mac}, based on the observation
that in lattice perturbation theory, a large fraction of renormalization 
effects comes from tadpole diagrams. The authors of \cite{lep_mac} propose to  
take the bulk of these effects into account by implementing a Mean 
Field Tadpole Improved (MFTI) prescription. MFT Improvement amounts to the 
following two substitutions:
\\ (1) The link variable
in any operator, such as $V_\mu^C$, is to be substituted by
\be
U_\mu \rightarrow \frac{U_\mu }{u_0}
\label{eq:u}
\ee
where $u_0$ is any reasonable mean-field estimate of the expectation
value of the link. Two standard choices are $u_0 = [\frac{1}{3} Tr U_P ]
^{1/4}$ and $u_0 = 1/(8K_c)$. In practice, for current values of the 
coupling constant, the two $u_0$ estimates differ by about $10\%$ (about 
$2\%$ for the Clover action). In the results of this work, the latter estimate
of $u_0$ has been used.\\ (2) The hopping parameter is obtained by 
substituting
\be
K \rightarrow \tilde K   = K u_0
\label{eq:K}
\ee
This implies, for example, that the MFTI quark mass is obtained by 
substituting
\be
am_0  \rightarrow a \tilde m =  \frac{1}{2\tilde K} - 4 
= 8 K_C (\frac{1}{2K} - \frac{1}{2K_C}) = 8K_C am
\label{eq:mtilde}
\ee
The ``standard" MFTI normalization of the quark field is given by the 
factor $\sqrt{2 \tilde K}$; its KLM/MFTI correction is obtained
through the substitution
\be
\sqrt{2K} F_\psi (am) \rightarrow
\sqrt{2 {\tilde K}} \tilde F_\psi (a\tilde m) = \sqrt{2 \tilde K 
(1+ a\tilde m)}
\ee
while for the conserved current $V^C_0$, the KLM/MFTI factor becomes
\be
F_{V_0^C} \rightarrow \tilde F_{V_0^C} = \frac{1}{u_0} \,\,
\left(1 + \frac{a\tilde m_1 + a \tilde m_2}{2} \right)^{-1}
\label{eq:fv0ctilde}
\ee
It must be kept in mind that the $1/u_0$ factor comes from the link $U_\mu$
present in the conserved current (i.e. it is a redefinition 
of the lattice operator due to the redefinition of the gauge field through 
eq.~(\ref{eq:u})) whereas the mass dependent factor arises from 
eq.~(\ref{eq:mtilde}) (i.e. it is a rescaling of the bare mass). This 
difference in the origin of the two factors will be important when we compute 
ME's of the conserved currents with KLM/MFTI corrections (see 
comments after eq.~(\ref{eq:r0mfti})). We also note that the prescription of 
eq.~(\ref{eq:mtilde}) has also been obtained by a tadpole resummation argument, 
without recourse to MFTI, in ref.~\cite{allmass}.

The KLM factors, having been calculated in the free theory and at zero quark 
spatial momentum, do not account for, say, $\calo(g_0^2 am)$,
$\calo(a\Lambda_{QCD})$ and $\calo(ap)$ corrections. The hope is that at 
large quark mass and small coupling, they account for most of the systematic 
error due to the finiteness of the UV cutoff. In sect. \ref{sec:npres} we 
will test these expectations.

\subsection{KLM Factors for the Clover Action}

We now calculate the KLM factors for the Clover case. Since
the free fermion propagator is the same for both the Wilson
and Clover actions, its KLM factor $F_\psi$ is the same.
However, fermion fields need to be rotated, as for example in
eq.~(\ref{eq:psir}). Since we usually opt for this specific
rotation ($z=1$), most correlation
functions are expressed in terms of the effective lattice propagator
$\cals^{eff} (x-y)$, given by eq.~(\ref{eq:srotz1}).
The KLM factor for the effective propagator is then given by
\be
\cals^{QCD}(t,\vecp = \vec 0) = a^{-3}
\frac{2K (1+am_0)}{ \left[1+\dfrac{1}{4} \left(1+am_0-\dfrac{1}{1+am_0}
\right) \right]^2} \, \cals^{eff} (t,\vecp = \vec 0)
\label{eq:cccc}
\ee
where again everything is worked out in the free case. Thus, compared to the 
Wilson case of eq.~(\ref{eq:klms}), there is an extra term, $F_R(am_0)$, arising 
from the fermion rotations in the improved KLM factor:
\be
F_\psi^I(am_0) = F_\psi(am_0) F_R(am_0) = (1 + am_0) F_R(am_0)
\ee
with
\be
F_R(am_0) = 
\left[1+\dfrac{1}{4} \left(1+am_0-\dfrac{1}{1+am_0} \right) \right]^{-2}
\label{eq:fr}
\ee
It can be seen immediately that the overall KLM factor, $F_\psi^I(am_0)$, 
is $1+\calo(am_0^2)$, in accordance with $\calo(a)$ improvement. From now 
on, we will directly substitute the bare mass $m_0$ by the subtracted one 
$m$, in the spirit of eq.~(\ref{eq:2kc}).

Let us consider the quark correlation function of the
local improved current $V_\mu^I = \bar \chi^R_1 \gamma_\mu 
\chi^R_2$ (i.e. the improved analogue of eq.~(\ref{eq:l})):
\be
L^I_\mu (t_x,t_y) = \sum_{\vec x ,\vec y} \langle
\chi^R_1(y) V_\mu^I(x) \bar \chi^R_2(0) \rangle 
\label{eq:li}
\ee
It is matched to its continuum counterpart through
\be
L^{QCD}_\mu = a^{-6} F_\psi^I(am_1) F_\psi^I(am_2) L^I_\mu(t_x,t_y)
\label{eq:vlclov}
\ee
Thus we see that also in the Clover case the KLM factor for the local 
vector current, as well as for any other local bilinear operator, is 1.

We now pass to the KLM factors of the conserved improved current of
eq.~(\ref{eq:vci}). Unlike the improved local current $V_\mu^I$,
fermion fields are not rotated for this current.
We first consider the improved quark correlation function of the current's
temporal component $V_0^{CI}$:
\be
C_0^I (t_y,t_x) = \sum_{\vec x ,\vec y} \langle
\chi^R _1(y) V_0^{CI}(x) \bar \chi^R _2(0) \rangle
\label{eq:clcorr}
\ee
The matching factors for this correlation are easily found to be:
\bea
\label{eq:clci0}
&& C_0^{QCD} (t_y,t_x) = \\
&& a^{-6} 2 K_1 2 K_2 F_\psi(am_1) F_\psi(am_2)
F_R^{1/2}(am_1) F_R^{1/2}(am_2)
F_{V_0^C} (am_1,am_2) C_0^I (t_y,t_x) \nonumber
\eea
where $F_{V_0^C}$ is given in eq.~(\ref{eq:fv0c}). Note that at $\vec p = \vec 
0$ the $\nabla _\mu \bar \chi \sigma_{\mu 0} \chi$ term of the current vanishes. 
The above overall KLM factor in eq.~(\ref{eq:clci0})
is $1 + \calo(a^2m^2)$, in accordance with improvement.

Finally, we examine the improved correlation $C^I_k (t_x,t_y)$ of the
spatial component of the conserved improved current:
\be
C_k^I (t_y,t_x) = \sum_{\vec x ,\vec y} \langle
\chi^R _1(y) V_k^{CI}(x) \bar \chi^R _2(0) \rangle
\label{eq:clcork}
\ee
The standard KLM computation yields the following matching:
\bea
\label{eq:clcik}
&& C_k^{QCD} (t_y,t_x) = \\
&& a^{-6} 2 K_1 2 K_2 F_\psi(am_1) F_\psi(am_2)
F_R^{1/2}(am_1) F_R^{1/2}(am_2)
F_{V_k^{CI}} (am_1,am_2) C_k^I (t_y,t_x)
\nonumber
\eea
where
\be
F_{V_k^{CI}} =
\left[ 1 - \frac{1 - (1+am_1)^2(1+am_2)^2}{4(1+am_1)(1+am_2)}\right]^{-1}
\label{eq:fvkci}
\ee
Again, being $1+\calo(am^2)$, the overall KLM factor in eq.~(\ref{eq:clcik})
is consistent with improvement.

We conclude by noting that using the Clover version of the axial WI, in the
spirit of KLM, requires a matching factor also for $2\rho$. From the
improved version of eq.~(\ref{eq:2rho}), with a symmetric lattice
temporal derivative $\overline \nabla^0$, we obtain the KLM matching
\be
2 \rho^{QCD} = \frac{2(am_1 + am_2)}
{\exp(am_1+am_2) - \exp(-am_1 - am_2)} 2\rho^I
\label{eq:2rhoklm}
\ee

Just like in the Wilson case, $\tilde m$'s and $u_0$'s can be introduced, 
whenever required by the prescription, in order to account for KLM/MFTI 
matching factors. Again, the extension of the Clover KLM factors to those 
characterizing hadronic correlation functions is straightforward.

\section{Non-perturbative Results for the RC's}
\label{sec:npres}

In this section we present results for the RC's $Z_V$ 
and $Z_A$ and the ratio $Z_S/Z_P$ obtained, non-perturbatively, 
from numerical studies of the chiral WI's on the lattice. The RC
of the local vector current, $Z_V$, has been also computed by taking
the ratio of ME's of the conserved to the local vector current
between different external states.

The main purpose of this study is to investigate the systematic errors present
in these calculations and, in particular, those due to the finiteness of the 
lattice cutoff. For this purpose, we have considered both the standard 
Wilson action and the tree-level $\calo (a)$-improved Clover action. In order to 
enhance the finite cutoff effects, we have studied a large range of values of 
the lattice bare quark masses, which vary from $am \simeq 0.05$ up to $am 
\simeq 1$. As expected, we find that these effects are typically larger with
the Wilson action than with the Clover action. 

Moreover, we have also tried to correct the 
finite cutoff effects by following the KLM prescriptions, discussed in the 
previous section. We find that KLM corrections are typically more effective 
in the Wilson case, where they are $\calo (am)$ rather than 
in the Clover case, where they are $\calo (a^2m^2)$.
The analysis of the Clover RC's
shows that the bulk of the finite cutoff effects
depends linearly on the bare quark mass. This
suggests the existence of large $\calo (g^2 am)$ terms, which,
being a consequence
of the quantum loops, cannot be corrected by the tree-level KLM 
prescriptions. 

A significant result is that the finite cutoff effects, even in the Clover 
case, can be as large as $20-30\%$, for values of the bare quark masses 
which are currently used in numerical simulations (e.g. $am \simeq 0.5$). 
This result strongly supports the necessity of an additional systematic 
improving strategy, such as \cite{lusch1,mrsstt,noi}. 
In the following we present our results in detail.

\subsection{Ratio Determination of $Z_V$ with the Wilson Action}
\label{subs:zvw}

We first present a computation of the RC of the local vector current, 
$Z_V$, in the case of the standard Wilson action. The data have been 
obtained by averaging over 20 quenched gauge field configurations, generated at
$\beta = 6.4$. This value of the lattice coupling constant corresponds to a
relatively large value of the lattice cutoff, $a^{-1} \simeq 3.6-3.8$ GeV. 
The lattice volume is $L^3 \times T = 24^3 \times 60$.
The same set of gauge configurations and quark 
propagators has already been used in previous phenomenological studies of
heavy meson physics on the lattice \cite{abada}.

$Z_V$ has been computed by considering different ratios of two- and 
three-point correlation functions which reduce, 
asymptotically in time, to different ME's of the conserved and the 
local vector currents; cf. eq.~(\ref{eq:rat}).
For simplicity, these ME's will be denoted
in the following as $<0 \vert V_k \vert \rho>$,
$<K \vert V_\mu \vert D>$ and $<K^* \vert V_\mu \vert D>$,
regardless of the specific flavour content of the mesons in the external
states. In particular, by $D$ we mean a pseudoscalar meson consisting of
a ``light" and a ``heavy" quark, whereas by $K$, $K^\ast$ and $\rho$ we
indicate mesons composed by two light quarks.
In all cases, the Wilson parameter of the light quark mass is fixed 
at $K = 0.1485$, the critical value being $K_C = 0.1506$. Thus this mass 
corresponds to a bare quark mass $am \simeq 0.05$. The heavy quark mass 
varies in the range between $am \simeq 0.2$ and $am \simeq 0.6$. 
In the three point correlation functions, the source of the final $K$ and
$K^*$ mesons is inserted at the time $t=0$. The $D$ meson is at $t=28$ and the 
vector current varies in the interval $12 \le t \le 16$. The $D$ meson
is always considered in its rest frame and the spatial momentum carried by the
current is denoted by $\vec q$. For simplicity,
the momentum transfers $\vec q = (0,0,0)$ and $\vec q = (1,0,0)$ (expressed
in lattice units of $2\pi/La$) will be abbreviated as
$q=0$ and $q=1$ respectively. 

A first way of obtaining $Z_V$ consists in 
computing the ratio $R_0$ of three point correlation functions defined in
eq.~(\ref{eq:rat}). This ratio has been computed at momentum transfer $q=0$.
In the large time limit, it is given by:
\be
R_0(q=0) = \frac{<K \vert V^C_0 \vert D>}{<K \vert V_0 \vert D>}
= Z_V + \calo (a)
\label{eq:r0}
\ee
According to the KLM claim, the bulk of the $\calo (a)$ corrections in this
expression should be given by the KLM factor $F_{V_0^C} (am_1,am_2)$ 
of eq.~(\ref{eq:fv0c}) associated with the vertex of the conserved current $V_0^C$ 
(the KLM factors $F_\psi(am)$, associated with the external quark fields,
cancel in $R_0$). Therefore we also define the ratio:
\be
R_0^{KLM} = R_0 \cdot F_{V_0^C} (am_1,am_2) =
 R_0 \cdot \left(1 + \frac{a m_1 + a m_2}{2}\right)^{-1} = Z_V + \calo (a)
\label{eq:r0klm}
\ee
If the KLM prescription works in this case, the $\calo (a)$ terms in
eq.~(\ref{eq:r0klm}) should be smaller than those in eq.~(\ref{eq:r0}). This
means that the ratio $R_0^{KLM}$
should exhibit a milder dependence on the two quark masses.
In addition, we also consider the MFTI version of the ratio $R_0$.
It is obtained from eq.~(\ref{eq:r0klm}), 
by substituting the bare quark masses with their MFTI expressions:
\be
R_0^{MFTI} = R_0 \cdot \left(1 + \frac{a \tilde m_1 + a \tilde m_2}{2}\right) 
^{-1} = Z_V + \calo (a)
\label{eq:r0mfti}
\ee
Notice that, with respect to the definition of the KLM/MFTI factor $\tilde
F_{V_0^C}$ given in eq.~(\ref{eq:fv0ctilde}), we have omitted in the previous 
equation a factor $u_0$. As pointed out after eq.~(\ref{eq:fv0ctilde}),
its presence would imply a redefinition
of the conserved current, whereas we are only interested in a comparison of
the discretization errors with and without KLM/MFTI corrections. Therefore,
by omitting $u_0$, we are computing the ratio $R_0^{MFTI}$ of the ME's of
the same operators which appear in the two ratios $R_0$ and $R_0^{KLM}$. 
In this way, all three ratios give estimates of the same $Z_V$ and only 
differ by $\calo (a)$ effects. 

\begin{figure}[t]
\ewxy{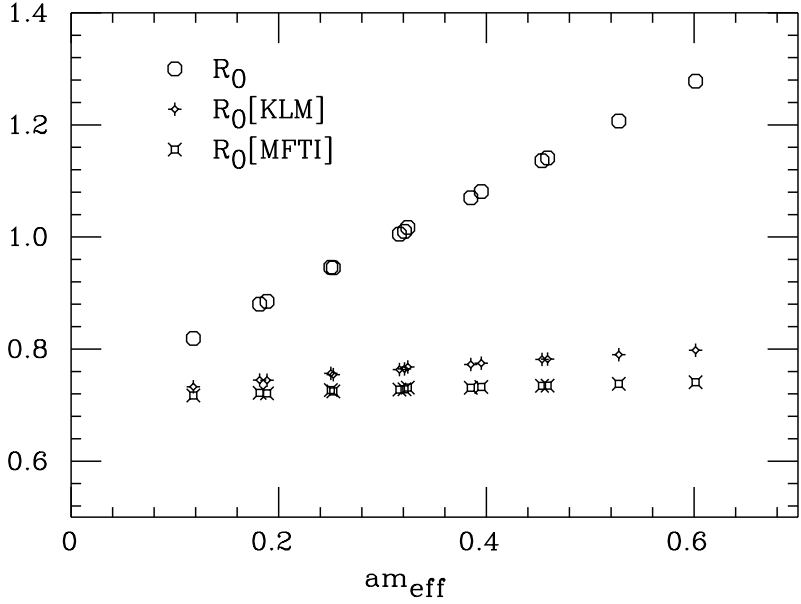}{100mm}
\vspace{3.0truecm}
\caption[]{\it The various ratios $R_0$, defined in eqs.~(\ref{eq:r0}),
(\ref{eq:r0klm})and (\ref{eq:r0mfti}) for Wilson fermions, as functions
of the effective quark mass.}
\protect\label{fig:fig1}
\end{figure}

The values of the three ratios $R_0$ are shown in table \ref{tab1},
for different values of the hopping parameter. They are also plotted
in figure \ref{fig:fig1} 
as a function of the ``effective" quark mass, $am_{eff} = (am_1+am_2)/2$.
\begin{table}
\centering
\begin{tabular}{|c|c|c|c|c|c|}\hline
$K_1$ & $K_2$ & $am_{eff}$ & $R_0$ & $R_0^{KLM}$  & $R_0^{MFTI}$  
\\ \hline
0.1275 &  0.1275 & 0.602 & 1.278 & 0.798 & 0.741 \\
0.1275 &  0.1325 & 0.528 & 1.207 & 0.790 & 0.738 \\
0.1275 &  0.1375 & 0.459 & 1.141 & 0.782 & 0.735 \\
0.1325 &  0.1325 & 0.454 & 1.136 & 0.782 & 0.735 \\
0.1275 &  0.1425 & 0.395 & 1.081 & 0.775 & 0.732 \\
0.1325 &  0.1375 & 0.385 & 1.070 & 0.773 & 0.731 \\
0.1275 &  0.1485 & 0.324 & 1.017 & 0.768 & 0.731 \\
0.1325 &  0.1425 & 0.321 & 1.010 & 0.764 & 0.728 \\
0.1375 &  0.1375 & 0.316 & 1.005 & 0.763 & 0.728 \\
0.1375 &  0.1425 & 0.253 & 0.945 & 0.754 & 0.725 \\
0.1325 &  0.1485 & 0.250 & 0.946 & 0.756 & 0.727 \\
0.1425 &  0.1425 & 0.189 & 0.885 & 0.744 & 0.721 \\
0.1375 &  0.1485 & 0.182 & 0.880 & 0.745 & 0.722 \\
0.1425 &  0.1485 & 0.118 & 0.820 & 0.733 & 0.718 \\
\hline
\end{tabular}
\caption{\it
The ratios $R_0$, $R_0^{KLM}$ and $R_0^{MFTI}$ for the Wilson action and
for different values of the quark masses. In the three point correlation 
functions the light spectator quark has mass $am \simeq 0.05$. The 
statistical errors in the table are less than
(or equal to) 1 on the last digit.  
}
\label{tab1}
\end{table}
We see that $R_0$ has a strong dependence on the bare quark
masses, varying by approximately $50\%$ as the mass varies from 
$am \simeq 0.2$ to $am \simeq 0.6$. This dependence is however significantly 
reduced when we apply the KLM correction. This reduction is somewhat
bigger in the MFTI prescription.

In order to discuss these results in a more quantitative way, we
fit the three ratios $R_0$, as a function of the quark masses,
taking into account contributions up to quadratic terms:
\be
R_0 = A + B (am_1 + am_2) + C (am_1 + am_2)^2 + D (am_1 - am_2)^2 
\label{eq:fitr0}
\ee
Since the current vertex is symmetric under interchange of the two
quark fields,
there is no linear dependence on the mass difference $(am_1-am_2)$.
For all three ratios $R_0$, $R_0^{KLM}$ and $R_0^{MFTI}$ the 
value of the coefficient $A$ in eq.~(\ref{eq:fitr0}) gives an estimate 
of $Z_V$, with presumably negligible $\calo (am)$ systematic errors,
which we denote by $Z_V^{eff}$. The result in all three cases is approximately 
the same and we quote, as an overall estimate:
\be
Z_V^{eff} = 0.71 (1) \qquad , \qquad {\rm{from}} \ \ R_0
\label{eq:zv_r0}
\ee
The parameter $B$ in eq.~(\ref{eq:fitr0}) gives the size of the leading $\calo
(am)$ corrections. We find that $B$ decreases from $B \simeq 0.46$
for $R_0$, to $B \simeq 0.09$ for $R_0^{KLM}$ and
to an even smaller value, $B \simeq 0.03$, for $R_0^{MFTI}$.
The coefficients $C$ and $D$ of the quadratic terms are always 
of the order $10^{-2}$ or smaller in magnitude.
Thus, in the case of the ratio $R_0$, the KLM 
prescription correctly takes into account and removes the bulk of the 
finite cutoff effects. 

Next we compare results for $Z_V$ obtained from different ratios of correlation 
functions, cf. eq.~(\ref{eq:rat}). In particular, for different values of the 
quark masses, we have computed: 
\bea
&& R_\rho (q=0) = 
\frac{\sum_r \epsilon^k_r \langle 0 \vert V^C_k (0) \vert \rho_r \rangle}
{\sum_r \epsilon^k_r \langle 0 \vert V_k (0) \vert \rho_r \rangle} = Z_V 
+ \calo (a) \nonumber \\
&& R_k (q=1) = 
\frac{\langle K \vert V^C_k (0) \vert D \rangle} {\langle K \vert V_k (0) 
\vert D  \rangle}  = Z_V + \calo (a) \\
&& R_k^\ast (q=1) = 
\frac{\sum_r \epsilon^\lambda_r \langle K^\ast_r \vert V^C_k (0) \vert D \rangle}
{\sum_r \epsilon^\lambda_r \langle K^\ast_r \vert V_k (0) \vert D \rangle} = 
Z_V + \calo (a) \nonumber 
\label{eq:rrr}
\eea
The KLM corrections to the correlation functions of these ratios 
always cancel out between numerator and denominator. Thus, according to
the KLM expectations,
all these ratios should be free of large $\calo (am)$ corrections. 

\begin{figure}[t]
\ewxy{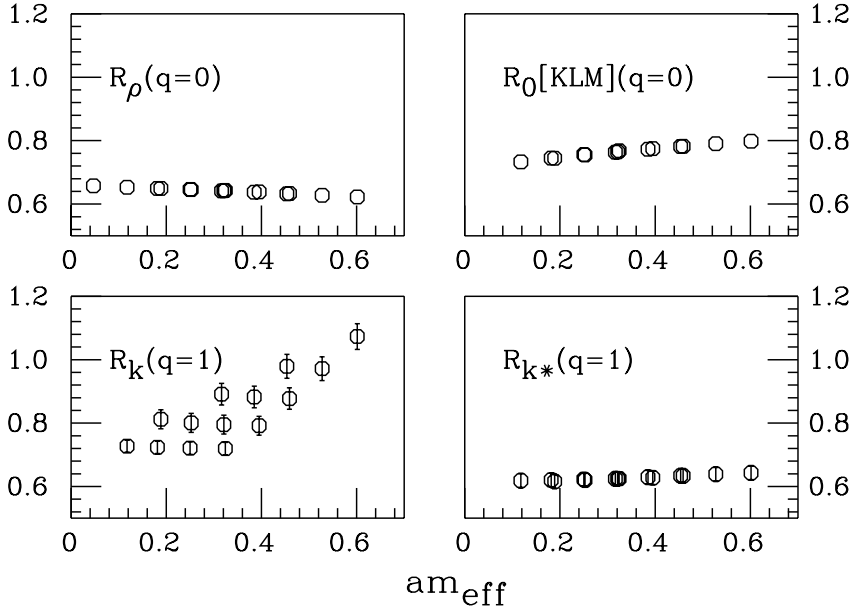}{100mm}
\vspace{3.0truecm}
\caption[]{\it The ratios $R_\rho$, $R_0^{KLM}$, $R_k$ and $R_k^\ast$, for 
Wilson fermions, as functions of the effective quark mass.}
\protect\label{fig:fig2}
\end{figure}

The values of $R_\rho$, $R_k$ and $R_k^\ast$ are given in table \ref{tab2} 
for different values of the vector current quark masses.
The same ratios are also 
plotted in figure \ref{fig:fig2} as a function of the effective quark mass. 
In the figure, the ratio $R_0^{KLM}$ is also plotted for comparison.
\begin{table}
\centering
\begin{tabular}{|c|c|c|c|c|c|}\hline
$K_1$ & $K_2$ & $am_{eff}$ & $R_\rho$ & $R_k$ & $R_k^\ast$ \\ \hline
0.1275  & 0.1275 &  0.602 &   0.622(1)  & 1.073(41) & 0.643(23) \\
0.1275  & 0.1325 &  0.528 &   0.628(1)  & 0.972(37) & 0.639(23) \\
0.1275  & 0.1375 &  0.459 &   0.633(1)  & 0.877(34) & 0.634(23) \\
0.1325  & 0.1325 &  0.454 &   0.632(1)  & 0.979(38) & 0.634(23) \\
0.1275  & 0.1425 &  0.395 &   0.638(1)  & 0.792(30) & 0.628(22) \\
0.1325  & 0.1375 &  0.385 &   0.637(1)  & 0.883(34) & 0.630(23) \\
0.1275  & 0.1485 &  0.324 &   0.643(2)  & 0.719(20) & 0.625(21) \\
0.1325  & 0.1425 &  0.321 &   0.642(1)  & 0.795(30) & 0.624(23) \\
0.1375  & 0.1375 &  0.316 &   0.641(1)  & 0.891(34) & 0.625(23) \\
0.1375  & 0.1425 &  0.253 &   0.645(1)  & 0.801(30) & 0.620(23) \\
0.1325  & 0.1485 &  0.250 &   0.646(2)  & 0.721(20) & 0.623(22) \\
0.1425  & 0.1425 &  0.189 &   0.649(1)  & 0.812(30) & 0.616(23) \\
0.1375  & 0.1485 &  0.182 &   0.649(2)  & 0.723(19) & 0.621(22) \\
0.1425  & 0.1485 &  0.118 &   0.653(2)  & 0.727(19) & 0.618(23) \\
0.1485  & 0.1485 &  0.047 &   0.658(2)  &  ---      &  ---      \\
\hline
\end{tabular}
\caption{\it
The ratios $R_\rho$, $R_k$ and $R_k^\ast$ for the Wilson action at
different values of the vector current quark masses.
In the three point correlation functions the light spectator quark has mass 
$am \simeq 0.05$.}
\label{tab2}
\end{table}
We see that $R_\rho$, $R_k$ and $R_k^\ast$ exhibit quite a
different behaviour as a function of the quark masses. When the mass varies 
from $am \simeq 0.2$ to $am \simeq 0.6$, the ratios $R_\rho$ and $R_k^\ast$ 
remain almost constant, while the ratio $R_k$ is found to be affected by large 
$\calo (a)$ corrections. In particular, by performing a fit as in
eq.~(\ref{eq:fitr0}), we find a large contribution coming
from the term proportional to $(am_1-am_2)^2$, with a coefficient
$D \simeq -0.51$. Compared to its values corresponding to the three
previous estimates of $R_0$, $D$ of $R_k$
is larger by approximately an order of magnitude.
The total finite cutoff effects
in the ratio $R_k$ can be as large as $50\%$ for $am \simeq 0.6$. These
effects cannot be corrected by the KLM prescription. 

The values of the coefficients $A$ of the fits give an estimate of $Z_V$
with presumably negligible $\calo (am)$ effects. We quote:
\bea
&& Z_V^{eff} = 0.66 (1) \qquad , \qquad {\rm{from}} \ \ R_\rho \nonumber \\
&& Z_V^{eff} = 0.68 (1) \qquad , \qquad {\rm{from}} \ \ R_k \\
&& Z_V^{eff} = 0.62 (1) \qquad , \qquad {\rm{from}} \ \ R_k^\ast \nonumber 
\eea
The differences between these values and the $R_0$ estimate of
eq.~(\ref{eq:zv_r0}) for $Z_V$ represent a measure of the
$\calo (a)$ corrections which remain after the $\calo 
(am)$ terms have been presumably removed by the fit.
These corrections, which are expected 
to be $\calo (aq)$ or $\calo (a \Lambda_{QCD})$, leave a systematic 
uncertainty which is approximately $15\%$.

\subsection{Ratio Determination of $Z_V$, $Z_A$ and $Z_P/Z_S$
with the Clover Action and Light Quark Masses}

In this subsection we present the results of a calculation of the RC's 
$Z_{V^I}$, $Z_{A^I}$ and the ratio $Z_{P^I}/Z_{S^I}$ obtained
with the tree-level $\calo (a)$-improved Clover action,
by using the lattice chiral WI's. From now on, the superscript ``$I$"
of the $Z_{O^I}$'s will be dropped.
We limit ourselves to light values of the bare quark masses.
The value of the lattice coupling constant is $\beta = 6.0$. The lattice
volume is $16^3 \times 32$. We have generated
18 quenched gauge field configurations.
We have worked at three values of the Wilson hopping 
parameters, namely $K = 0.1410$, $0.1425$ and $0.1432$ corresponding
to bare quark masses ranging from $am=0.056$ to $am=0.111$.
The critical value is $K_C=0.14554$.

All two- and three-point correlation functions have been computed by 
considering degenerate quark masses. The results are a replication
of those of ref.~\cite{clvwi} for three quark masses. The RC of the local 
axial current $Z_A$ and the ratio $Z_P/Z_S$ have been obtained from
gauge invariant chiral WI's, between external hadron states (see
eqs.~(\ref{eq:pav}) and (\ref{eq:wisp})).
Gauge dependent chiral WI's between external quark states have also been
implemented for a calculation of $Z_A$ (see eq.~(\ref{eq:zaqb})).
The RC of the local vector current, $Z_V$, has been 
computed from the ratio $R_\rho$, with zero spatial momenta.
The results of this computation are presented 
in table \ref{tab3}.
\begin{table}
\centering
\begin{tabular}{|c|c|c|c|c|c|}\hline
$K_l$ & $am_l$ & $Z_V$ & $Z_A$ (g.i.) & $Z_A$ (g.d.)  & $Z_P/Z_S$ \\ \hline
0.1410 & 0.111 & 0.833(1) & 1.06(5) & 1.07(7) & 0.61(3) \\
0.1425 & 0.073 & 0.824(2) & 1.09(3) & 1.14(8) & 0.60(2) \\
0.1432 & 0.056 & 0.819(2) & 1.07(7) & 1.17(13)& 0.60(3) \\
\hline
\end{tabular}
\caption{\it
The RC's $Z_V$, $Z_A$ and the ratio $Z_S/Z_P$ for the Clover action with light 
quark masses. The labels ``g.i." and ``g.d" distinguish results obtained from
gauge-invariant and gauge-dependent WI's respectively.}
\label{tab3}
\end{table}

We see that, in the case of $Z_V$, the statistical errors are sufficiently small 
to reveal the presence of finite cutoff effects in the calculation. By performing 
a fit as in
eq.~(\ref{eq:fitr0}) we can obtain the value of $Z_V$, with presumably
negligible $am$ effects. We quote:
\be
Z_V^{eff} = 0.80 (1) \qquad , \qquad {\rm{from}} \ \ R_\rho
\label{eq:zv_clo_1}
\ee
As seen in table \ref{tab3}, the variations of
the RC $Z_A$ and the ratio $Z_S/Z_P$
with the light quark masses are of the size of the statistical errors.
Thus, in these cases the presence of finite cutoff effects cannot be
determined. An analogous study at $\beta = 6.2$ has been performed
by the UKQCD collaboration \cite{ukzs}.

\subsection{Ratio Determination of $Z_V$ with the Clover Action and 
Heavy Quark Masses}
\label{sub6_3}

In order to investigate the size of finite cutoff effects when the
tree-level $\calo (a)$-improved Clover action is used, we have also 
computed $Z_V$ at several large values of the quark mass. We have generated
30 quenched gauge field configurations at $\beta = 6.0$. The lattice volume 
is $16^3 \times 48$. Two different values of the light quark hopping parameter
have been considered ($K=0.1425$ and $K=0.1432$) and four different values 
of the heavy quark mass. The latter correspond to bare quark masses varying 
from $am \simeq 0.3$ up to $am \simeq 0.9$.

We have computed $Z_V$ from the ratios $R_0$ and
$R_\rho$ of eq.~(\ref{eq:rat}), with quark propagators improved according to
eq.~(\ref{eq:srotz1}).
All external states and currents
are always at zero spatial momentum. In the three-point correlation 
functions the pseudoscalar $D$ meson is inserted in the middle of the lattice, 
at $t = 24$, so that results from both positive and negative times can be 
averaged, thus improving the statistics. 

In order to investigate the effects of the KLM corrections in the Clover case,
we have also considered the corresponding KLM-improved versions
of the ratios $R_0$ and $R_\rho$. According to eqs.~(\ref{eq:vlclov}), 
(\ref{eq:clci0}) and (\ref{eq:clcik}), these corrections must be defined as: 
\bea 
&& R_0^{KLM} = \frac{F_R(am_1)^{1/2}F_R(am_2)^{1/2}}{F_{V_0^C}(am_1,am_2)} 
\cdot R_0 = Z_V + \calo (g^2a,a^2) \nonumber \\ 
&& R_\rho^{KLM} = \frac{F_R(am_1)^{1/2}F_R(am_2)^{1/2}}{F_{V_k^{CI}}(am_1,am_2)} 
\cdot R_\rho = Z_V + \calo (g^2a,a^2)
\label{eq:rcloklm}
\eea
where the several KLM factors are given in eqs.~(\ref{eq:fv0c}),
(\ref{eq:fr}) and (\ref{eq:fvkci}). Notice that, in the Clover case,
the KLM corrections are always of $\calo (a^2)$.

\begin{figure}[t]
\ewxy{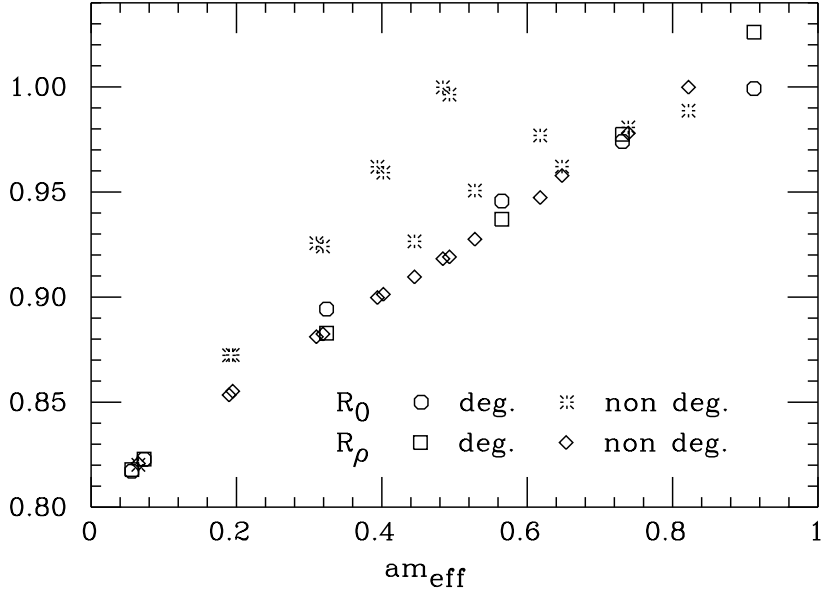}{100mm}
\vspace{3.0truecm}
\caption[]{\it The ratios $R_0$ and $R_\rho$, for Clover fermions, as a 
function of the effective quark mass. We distinguish between the cases
of degenerate and non-degenerate quarks in the current.}
\protect\label{fig:fig3}
\end{figure}

\begin{figure}[t]
\ewxy{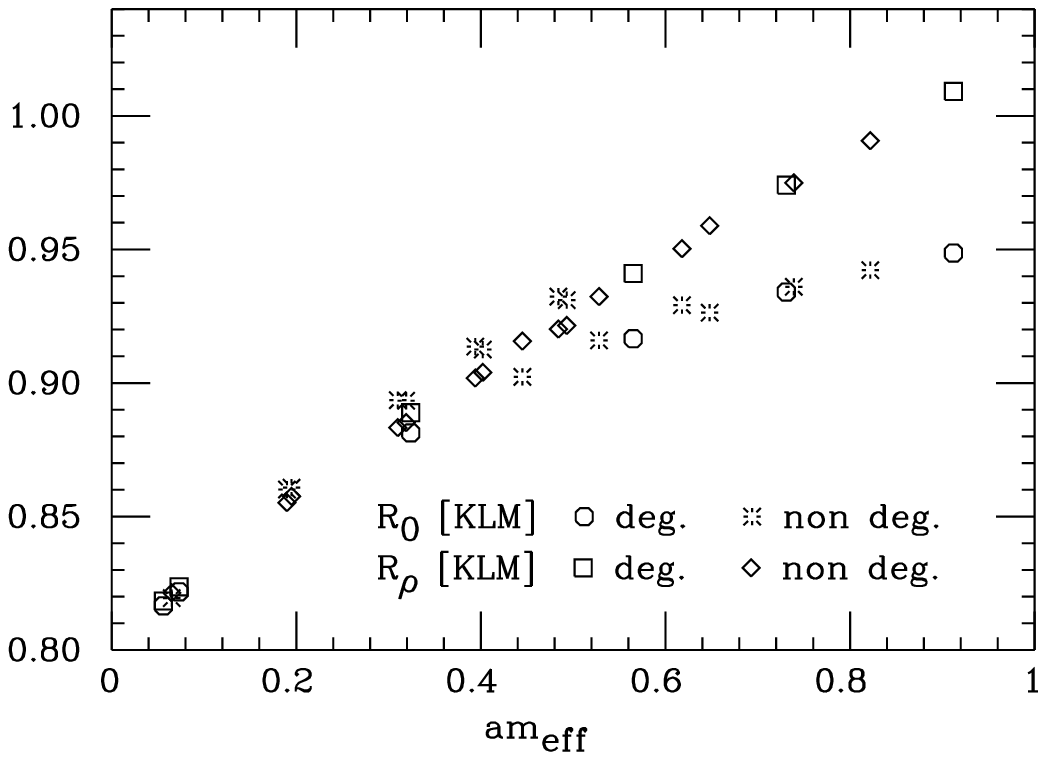}{100mm}
\vspace{3.0truecm}
\caption[]{\it Same as in figure 3 with KLM corrections.}
\protect\label{fig:fig4}
\end{figure}

The values of $R_0$, $R_\rho$ and their KLM counterparts
are given in table \ref{tab4} for several values of the quark 
masses. The same ratios are also plotted in figures \ref{fig:fig3} and 
\ref{fig:fig4}, with and without the KLM corrections, as a function of the 
effective mass $am_{eff} = (am_1 + am_2)/2$. 
\begin{table}
\centering
\begin{tabular}{|c|c|c|c|c|c|c|}\hline
$K_1$ & $K_2$ & $am_{eff}$ & $R_0$ & $R_0^{KLM}$ & $R_\rho$ & $R_\rho^{KLM}$
\\ \hline
   0.1150 & 0.1150 & 0.912 & 0.999 & 0.949 & 1.026 & 1.009 \\
   0.1150 & 0.1200 & 0.822 & 0.989 & 0.942 & 1.000 & 0.991 \\
   0.1150 & 0.1250 & 0.739 & 0.981 & 0.936 & 0.978 & 0.976 \\
   0.1200 & 0.1200 & 0.731 & 0.974 & 0.934 & 0.977 & 0.974 \\
   0.1200 & 0.1250 & 0.648 & 0.962 & 0.926 & 0.958 & 0.959 \\
   0.1150 & 0.1330 & 0.618 & 0.977 & 0.929 & 0.947 & 0.950 \\
   0.1250 & 0.1250 & 0.565 & 0.946 & 0.917 & 0.937 & 0.941 \\
   0.1200 & 0.1330 & 0.528 & 0.951 & 0.916 & 0.927 & 0.932 \\
   0.1150 & 0.1425 & 0.493 & 0.996 & 0.931 & 0.919 & 0.922 \\
   0.1150 & 0.1432 & 0.484 & 0.999 & 0.932 & 0.918 & 0.920 \\
   0.1250 & 0.1330 & 0.445 & 0.927 & 0.902 & 0.910 & 0.916 \\
   0.1200 & 0.1425 & 0.402 & 0.959 & 0.913 & 0.901 & 0.903 \\
   0.1200 & 0.1432 & 0.394 & 0.962 & 0.914 & 0.899 & 0.902 \\
   0.1330 & 0.1330 & 0.324 & 0.894 & 0.881 & 0.883 & 0.889 \\
   0.1250 & 0.1425 & 0.319 & 0.924 & 0.893 & 0.882 & 0.885 \\
   0.1250 & 0.1432 & 0.310 & 0.926 & 0.894 & 0.881 & 0.883 \\
   0.1330 & 0.1425 & 0.195 & 0.872 & 0.861 & 0.855 & 0.858 \\
   0.1330 & 0.1432 & 0.190 & 0.872 & 0.860 & 0.853 & 0.855 \\
   0.1425 & 0.1425 & 0.073 & 0.823 & 0.822 & 0.823 & 0.824 \\
   0.1425 & 0.1432 & 0.065 & 0.820 & 0.819 & 0.821 & 0.821 \\
   0.1432 & 0.1432 & 0.056 & 0.817 & 0.817 & 0.818 & 0.818 \\
\hline
\end{tabular}
\caption{\it
The ratios $R_0$ and $R_\rho$, and their KLM-improved counterparts,
for the Clover action at different values of the quark masses. The 
statistical errors are always less than (or equal to) 1
on the last digit.} 
\label{tab4}
\end{table}
The main feature shown in figure \ref{fig:fig3} is that both $R_0$ and $R_\rho$ 
receive 
a large contribution from a term which is linear in the quark mass. This term 
affects the value of the two ratios by approximately $15\%$ at $am \simeq 0.5$ 
and by about $25\%$ at $am \simeq 0.9$.
Moreover, such a large systematic error 
cannot be corrected by the KLM factors, the latter being at least quadratic in 
the quark masses. In fact, a qualitatively similar behaviour is found for the 
points in figure \ref{fig:fig4}, where the KLM correction has been implemented.

By looking at figure \ref{fig:fig3} we also notice that the values of $R_0$,
obtained with degenerate quark masses of the vector current,
and those obtained by non-degenerate quarks,
do not lie on the same curve. This suggests a 
significant dependence of this ratio on terms quadratic in the mass difference
$(am_1 - am_2)^2$. 
In this respect,
figure \ref{fig:fig4} shows that the KLM correction improves the situation,
since the
difference between degenerate and non-degenerate points is largely reduced. 
However, the KLM prescription cannot correct large $\calo(g^2am)$
effects, which are evidently present in figure \ref{fig:fig4}. It would
also appear from this figure that such terms are independent of the ME's, 
since the difference between $R_0$ and $R_\rho$ settles only after masses of 
approximately $am \simeq 0.5$.

To make this discussion more quantitative, we fit the
ratios $R_0$, $R_\rho$ and their KLM counterparts, as
in eq.~(\ref{eq:fitr0}). The resulting
values of the coefficients $A$, $B$, $C$ and $D$ are given in table 
\ref{tabcoe}.
\begin{table}
\centering
\begin{tabular}{|l|c c c c|}\hline
               & $A$  & $B$  &   $C$   & $D$    \\ \hline
$R_0$          & 0.80 & 0.19 & -0.05   & 0.09   \\
$R_0^{KLM}$    & 0.80 & 0.16 & -0.05   & 0.03   \\ \hline
$R_\rho$       & 0.81 & 0.12 & -0.0008 &-0.002  \\
$R_\rho^{KLM}$ & 0.80 & 0.14 & -0.02   &-0.008  \\ \hline
\end{tabular}
\caption{\it
Values of the coefficients $A$, $B$, $C$ and $D$ obtained from a fit
of the ratios $R_0$, $R_\rho$ and their KLM-improved counterparts.} 
\label{tabcoe}
\end{table}
The values of the coefficient $A$, in table \ref{tabcoe}, represent an estimate 
of $Z_V$ which is presumably free from any $\calo (am)$ corrections.
We then quote:
\be
Z_V^{eff} = 0.80 (1) \qquad , \qquad {\rm {from}} \ \ R_0  \ , \ R_\rho
\label{eq:zv_clo_2}
\ee
in remarkable agreement with its determination from light quark masses;
cf. eq.~(\ref{eq:zv_clo_1}). Notice that the two results have been obtained from
completely independent numerical simulations.

The values of the coefficients $B$ in table \ref{tabcoe} provide
an estimate of the size of the
correction coming from a term which is linear in the quark mass. Since the 
action is tree-level improved, this term must be of $\calo (g^2a)$ or higher. 
We find $B\sim 0.15$, for both ratios $R_0$ and $R_\rho$. As expected, these 
coefficients are practically unaffected by the KLM corrections. Remarkably, 
however, we find that their value is almost ``universal'', in the sense that it
is almost independent on the external states of the ME of the vector current. 

On the other hand,
the values of the coefficients $C$ and $D$ of the quadratic terms 
are not universal. For the ratio $R_0$,
the large value of coefficient $D$ is found to be significantly
reduced by the KLM correction.

We stress again that all these $\calo (a)$ effects are quite relevant
in the region which, for current values of the lattice spacing, corresponds
approximately to the charm quark mass. This in turn is crucial for the 
phenomenological studies of the heavy flavour physics on the lattice.

In an attempt to reduce the large $\calo (g^2a)$ effects observed in the
data, we have also tried to modify the definition of the improved operators 
by varying the $z$ parameter in eq.~(\ref{eq:ipropz}). Specifically,
we have considered the values $z=0$ and $z=0.5$ besides the standard choice 
$z=1$ analyzed so far. We find that in all cases the size of the $\calo 
(g^2a)$ contributions is approximately the same. Thus, no specific choice of 
$z$ appears to improve the situation.
 
\subsection{WI Determination of $Z_V$ and $Z_A$
with the Clover Action and Heavy Quark Masses}

In this subsection we present the Clover results for the RC's $Z_V$ and 
$Z_A$ obtained from a direct evaluation of the axial lattice chiral WI. 
This calculation has been performed by using the set of gauge configurations 
and quark propagators of subsection \ref{sub6_3}.

A determination of $Z_V$ can be obtained by considering the axial 
WI of eq.~(\ref{eq:vuds}). In the Clover case, the right-hand side of this 
equation must be corrected for an $\calo (a)$ term as shown in
eq.~(\ref{eq:audsi}), so that the resulting WI has the form:
\be
2 \rho^{12} \int dx \int d \vec y \langle P^{12}(x) A^{21}_0 (y)
V^{22}_0 (0) 
\rangle = - (\frac{1}{Z_V^{22}} - a \rho^{12})
\int d \vec y \langle A^{21}_0 (y) A^{12}_0 (0) \rangle
\label{eq:audsi00}
\ee
By computing $\rho$ from eq.~(\ref{eq:2rho}) (with improved operators and symmetric
derivative $\overline{\nabla}^0$) and the two correlation functions appearing
in the above WI, we can solve for $Z_V$.

The KLM improvement to eq.~(\ref{eq:audsi00}) involves both the standard KLM 
factors, which correct the two- and three-point correlation functions, and
the KLM correction to the ratio $\rho$, given by eq.~(\ref{eq:2rhoklm}). 
The results for $Z_V$ obtained from this WI are then presented
in table \ref{tab5}, 
with and without the KLM corrections. For a comparison, we also show in the table 
the values of $Z_V$ corresponding to the ratio $R_0^{KLM}$, which have
already been given in table \ref{tab4}.
\begin{table}
\centering
\begin{tabular}{|c|c|c|c|c|c|}\hline
$K_1$ & $K_2$ & $am_{eff}$ & $Z_V$ & $Z_V^{KLM}$ & $Z_V \ (R_0^{KLM})$ \\ \hline
  0.1150 & 0.1150 & 0.912 & 0.62(1) & 1.01(1) & 0.949 \\
  0.1200 & 0.1200 & 0.731 & 0.68(1) & 0.93(1) & 0.934 \\
  0.1250 & 0.1250 & 0.565 & 0.74(1) & 0.88(1) & 0.917 \\
  0.1330 & 0.1330 & 0.324 & 0.79(1) & 0.84(1) & 0.881 \\
  0.1425 & 0.1425 & 0.073 & 0.80(2) & 0.80(2) & 0.822 \\
  0.1432 & 0.1432 & 0.056 & 0.79(2) & 0.79(2) & 0.817 \\
\hline
\end{tabular}
\caption{\it
The RC $Z_V$, for the Clover action, obtained from the axial 
WI with and without the KLM correction. For comparison, the values of the ratio 
$R_0^{KLM}$ are also shown in the last column.}
\label{tab5}
\end{table}

Let us first consider in table \ref{tab5} the values of $Z_V$ obtained from the WI
(\ref{eq:audsi00}) without the KLM corrections. As the bare quark mass increases 
from $am \simeq 0$ up to $am \simeq 0.9$ we find that $Z_V$ decreases
by approximately
$25\%$ due to finite cutoff effects. This is to be contrasted to the
increase by about $25\%$ of the $R_0$ estimate of $Z_V$ 
(sixth column of the table) in the same mass interval.
The two determinations only agree at very small values of the quark mass. 

When we apply the KLM prescription we see from the fifth column of
table \ref{tab5} a dramatic change in the estimates of $Z_V$.
This effect is mainly due to the different normalization of the ratio
$\rho$, and signals the presence of large $\calo (a^2m^2)$ corrections.
However, in spite of the KLM correction,
the variation of $Z_V$ with the quark mass is still of about $25\%$.
Comparing the results of the last two columns of the table, we note
a good agreement between the KLM corrected $Z_V$ estimates obtained with
the WI and Ratio methods. This suggests that the KLM correction 
has adjusted the
$\calo (a^2)$ terms in such a way that the resulting values of the
RC's exhibit a more universal behaviour, approximately 
independent of the specific correlation functions used for their computation.

Fitting $Z_V$ as in eq.~(\ref{eq:fitr0}), we obtain
an estimate of the RC's with presumably negligible $\calo (am)$ 
discretization effects. We quote:
\be
Z_V^{eff} = 0.80 (2) \qquad , \qquad {\rm {from \ \ WI} }
\label{eq:zv_clo_3}
\ee
in perfect agreement with the previous determinations from the Ratio method
(eqs.~(\ref{eq:zv_clo_1}) and (\ref{eq:zv_clo_2})).

\begin{figure}[t]
\ewxy{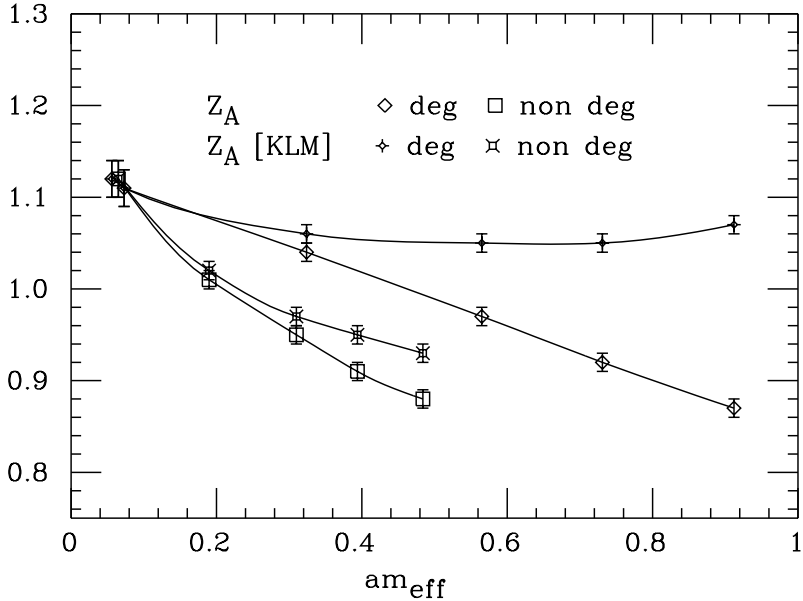}{100mm}
\vspace{3.0truecm}
\caption[]{\it 
The RC $Z_A$ for the Clover action, with and without 
KLM corrections, as a function of the effective quark mass. Lines are
only guides to the eye.}
\protect\label{fig:fig5}
\end{figure}

The RC of the local axial current $Z_A$, can be computed from the WI of 
eq.~(\ref{eq:audsi}), by choosing suitable spatial Lorentz indices of the 
currents: $\nu = \rho = 1,2,3$. Also in this case we find it convenient to 
take the flavour -2 and -3 quarks degenerate in mass, so that the WI only 
involves the RC's $Z_A^{12}$ and $Z_V^{22}$. For the latter, we use the 
values obtained from the ratio $R_0$ (at the corresponding quark masses), 
which is affected by a smaller statistical error, compared to the 
determination coming from the axial WI. Also for $Z_A$ we have considered 
the results obtained with and without the KLM prescription, which involves
in this case corrections to the correlation functions, to the ratio $\rho$ 
and to $Z_V$ itself. The resulting values of $Z_A$ are presented in table 
\ref{tab6} and are plotted in figure \ref{fig:fig5} as a function of the 
effective mass $am_{eff} = (am_1 + am_2)/2$.
\begin{table}
\centering
\begin{tabular}{|c|c|c|c|c|}\hline
$K_1$ & $K_2$ & $am_{eff}$ & $Z_A$ & $Z_A^{KLM}$ 
\\ \hline
0.1150 &  0.1150 & 0.912 &  0.87(1) &    1.07(1)  \\
0.1200 &  0.1200 & 0.731 &  0.92(1) &    1.05(1)  \\
0.1250 &  0.1250 & 0.565 &  0.97(1) &    1.05(1)  \\
0.1150 &  0.1432 & 0.484 &  0.88(1) &    0.93(1)  \\
0.1200 &  0.1432 & 0.394 &  0.91(1) &    0.95(1)  \\
0.1330 &  0.1330 & 0.324 &  1.04(1) &    1.06(1)  \\
0.1250 &  0.1432 & 0.310 &  0.95(1) &    0.97(1)  \\
0.1330 &  0.1432 & 0.190 &  1.01(1) &    1.02(1)  \\
0.1425 &  0.1425 & 0.073 &  1.11(2) &    1.11(2)  \\
0.1425 &  0.1432 & 0.065 &  1.12(2) &    1.12(2)  \\
0.1432 &  0.1432 & 0.056 &  1.12(2) &    1.12(2)  \\
\hline
\end{tabular}
\caption{\it
The RC $Z_A$ for the Clover action, with and without 
KLM corrections, for different values of the quark masses.}
\label{tab6}
\end{table}

As in the case of the vector current, we find from table \ref{tab6} and figure 
\ref{fig:fig5} a large dependence of $Z_A$ on the two quark masses. For values
of the effective quark mass of the order of $am \simeq 0.5$ such a dependence can
affect the value of the RC by approximately $20\%$. When we fit $Z_A$ as in
eq.~(\ref{eq:fitr0}), we find for the coefficient of the linear term the value
$B \simeq -0.13$. This value is of opposite sign but similar in magnitude to the
coefficient found for $Z_V$. This indicates the presence of large 
$\calo (g^2am)$ corrections in both cases. From figure \ref{fig:fig5} we also
notice that points corresponding to degenerate and non degenerate quark masses
in the current do not lie on the same curve, indicating the presence
of a large dependence on the difference of the two quark masses. 
From the fit 
we obtain $D \simeq -0.17$ for both uncorrected and KLM corrected data.
Therefore, we find that the KLM corrections do not improve the 
finite cutoff effects observed in the numerical estimates of $Z_A$.
 
From the zeroth order coefficient in the fit of $Z_A$ we also derive an 
estimate of the value of the RC with presumably negligible $\calo (am)$ 
effects. We quote:
\be
Z_A^{eff} = 1.10 (2)
\label{eq:za_clo}
\ee
in good agreement with the independent determination at light quark masses,
presented in table \ref{tab3}.

\section{Further Improvements and Future Outlook}
\label{sec:conc}

The results presented in this paper lead to the following conclusions:
\begin{itemize}
\item{
Applying various recipes of BPT gives RC's which are consistent to each
other within $10\%$ and which differ from those obtained from SPT by at most
$20\%$. In most (but not all) cases, perturbative improvements based on BPT 
can lead to more reliable estimates of the RC's, in the sense that they
are in better agreement with those obtained non-perturbatively.}
\item{
At present values of the lattice cutoff, non-perturbative estimates of the 
RC's based on lattice WI's, suffer, in the Wilson case, from large
discretization errors of about $20\%$. When tree-level Clover improvement is
implemented, such systematic effects decrease significantly for correlation 
functions computed at small quark mass. They remain large, however,
as the quark mass increases ($am \simge 0.2$). We have checked that this
situation persists for various choices of the field rotation, characteristic
of tree-level $\calo (a)$ improvement (i.e. different choices
of $z$ in eq.~(\ref{eq:ipropz})).}
\item{
A very systematic analysis of the KLM corrections has been carried out, both
with Wilson and Clover actions. In the Wilson case, the KLM factors sometimes
succeed and sometimes fail to correct large $\calo (am)$ effects, depending
on the correlation function examined. Moreover, even in the range of small
quark masses, large residual $\calo (a)$ errors remain, as can be seen from the 
discrepancies in the $Z_V$ estimates from different correlation functions.
In the Clover case, KLM cannot correct the observed large $\calo (g_0^2 am)$
systematic errors.}
\end{itemize}

The implication of these results is that, with present day lattice
cutoffs of about $2 - 4$ GeV ($\beta = 6.0 - 6.4$), simulations
of QCD phenomenology in the charm region (e.g. studies of $f_D$ or $D$-meson
semileptonic decays) suffer from sizeable finite cutoff errors. This
limits the predictability of lattice QCD for heavy flavour physics. 
In order to circumvent this problem, and given the impracticability of 
performing the simulations at significantly smaller lattice spacings, one must 
seek to improve the lattice convergence to the continuum limit. In this
work, we have examined two such efforts, namely Clover tree-level
improvement and KLM matching. Here we comment briefly on the recently
proposed extensions of these efforts, \cite{lusch1} and \cite{ekm}.

The aim of ref.~\cite{lusch1} is non-perturbative $\calo (a)$ improvement. 
To achieve this, the Wilson action must be modified by adding the (dimension-5) 
Clover term of eq.~(\ref{eq:sclov}) with the coefficient $c_{SW}$ determined
non-perturbatively. In addition, a dimension-$d$ renormalized operator must 
be cast in the form
\be
\hat O = Z_O(g^2,a\mu) [1 + b_O(g^2) a m] [O + a \sum_n c_n( g^2) \delta O_n]
\label{eq:climl} 
\ee
where the $\delta O_n$'s are the dimension-$(d+1)$ operators which are allowed 
to mix with $O$. 
The constants $b_O$ and $c_n$ can either be determined 
in PT or non-perturbatively. In PT they are known at tree-level and, in some cases,
at 1-loop \cite{lw}. For example, for bilinear local quark operators, 
$c_n = 0$ and $b_0 = 1$ correspond to the tree-level Clover improvement 
discussed in this work, with field rotations determined at $z=1$. 
For a full $\calo (a)$ improvement, all the improving coefficients must
be computed non-perturbatively. In ref.~\cite{lusch1} the coefficients $c_{SW}$ 
of the Clover term and
$c_A$ of the axial current have been computed non-perturbatively, with the 
aid of lattice chiral WI's; for the coefficient $c_V$ of the vector current see
ref.~\cite{guagn}. As shown in \cite{mrsstt}, chiral WI's cannot be used in the
spirit of \cite{lusch1} for the determination of the coefficients $b_O$ (with
the exception of $b_V$ of the vector current computed in \cite{lusch}). Two 
methods for the computation of $c_O$ and $b_O$, which achieve $\calo (a)$ improvement 
at non-zero quark masses, can be found in refs.~\cite{mrsstt} and \cite{noi}.
In this way, the approach to the continuum limit is accelerated to $\calo (a^2)$.

The proposal of ref.~\cite{ekm} can be summarized as follows. We saw that 
the KLM matching consists in the redefinition
\be
\hat O = Z_O(g_0^2,a\mu) F_O(am) O
\ee
The KLM factor $F_O$ has been worked out at zero spatial momentum in the 
free theory and for any value of the quark mass. The authors of \cite{ekm}
propose to extend this calculation beyond the free theory. They claim that 
it is possible to perform the lattice to continuum matching, again at zero 
spatial momentum but for any value of the quark mass, order by order in PT
and to all orders in $am$. However, this matching has not yet been worked
out beyond tree-level. 

We believe that the non-perturbative Clover improvement of on-shell ME's,
proposed in \cite{lusch1}, is the most promising way of controlling 
discretization errors at present day values of the UV cutoff. Our analysis 
of the tree-level Clover improvement at large quark masses strongly suggests 
that it is crucial to implement the extension of this program beyond the 
chiral limit \cite{mrsstt,noi}. This will hopefully render lattice computations 
of heavy flavour physics extremely accurate.

\section *{Acknowledgements}
We wish to thank C.R. Allton, G.Martinelli and S.Petrarca for their early 
participation in this work. We are very grateful to the above and also to
G.C Rossi, M.Talevi and M.Testa for their constant help and many suggestions. 
We acknowledge useful discussions with A.Kronfeld and P.Mackenzie. We thank 
the members of the ELC collaboration for providing us with their data, which 
we have used in the $\beta=6.4$ numerical computation with Wilson fermions.

\section*{Appendix: The Calculation of $q^\ast$}

We give here the essential details and discuss some subtleties concerning
the calculation of $q^\ast$ for the RC's $Z_V$, $Z_A$ and the ratio $Z_S/Z_P$.
For definitiveness, we consider the $\msb$ renormalization prescription in the 
$NDR$ scheme. The general 1-loop expression of eq.~(\ref{eq:rlatZ}), for the
three quantities of interest, becomes
\bea
Z_V &=& 1 - \frac{g_0(a)^2}{(4\pi)^2} C_F \left[ C_V^{LAT} + C_\Sigma^{LAT} 
\right]
\nonumber \\
Z_A &=& 1 - \frac{g_0(a)^2}{(4\pi)^2} C_F \left[ C_A^{LAT} + C_\Sigma^{LAT} 
\right]
\label{eq:zvasp} \\
\frac{Z_S}{Z_P} &=& 1 - \frac{g_0(a)^2}{(4\pi)^2} C_F \left[ C_S^{LAT} - 
C_P^{LAT} \right]
\nonumber
\eea
These expressions are easily derived from the discussion preceeding
subsec.\ref{sec:bpt}.
For convenience, we have rescaled the lattice constants
$C^{LAT}$ by the colour factor $C_F=4/3$, which now appears explicitly. 
Note that none of the above quantities depends on the
renormalization scale $a\mu$. Consequently, the $q^\ast$'s of these quantities
are also independent of $a\mu$.

For the Wilson action, the terms on the r.h.s of the above expressions have 
been worked out in \cite{mz,gaym}:
\bea
C^{LAT}_V + C^{LAT}_\Sigma &=&
1 - \left[ \frac{1}{4} I_1 - \frac{1}{2} I_2 - I_3 + I_\Sigma \right]
\nonumber \\
C^{LAT}_A + C^{LAT}_\Sigma &=&
1 - \left[ \frac{1}{4} I_1 + \frac{1}{2} I_2 - I_3 + I_\Sigma \right]
\label{eq:Is} \\
C^{LAT}_S - C^{LAT}_P &=& - 2 I_2 \nonumber
\eea
where the finite integrals $I_1$, $I_2$, $I_3$ and $I_\Sigma$ are
reproduced here for completeness:
\bea
&& I_1 = 4 \pi^2 \int_{-\pi}^{\pi} \frac{d^4q}{(2\pi)^4} \left[
\frac{ \frac{4}{3}(\Delta_1 \Delta_4 - \Delta_5) + 2 r^2 \Delta_4^2
+ 4r^4 \Delta_1^2 \Delta_4 }{\Delta_1 \Delta_2^2}
+ \frac{4 \Delta_4 (\Delta_3^2 -2 \Delta_1 \Delta_3)}{\Delta_1^3 \Delta_2^2} \right]
\nonumber \\
&& I_2 = 4 \pi^2 r^2 \int_{-\pi}^{\pi} \frac{d^4q}{(2\pi)^4}
\left[ \frac{4 \Delta_1 (4-\Delta_1) - \Delta_4}{\Delta_2^2} \right]
\\
&& I_3 = 4 \pi^2 \int_{-\pi}^{\pi} \frac{d^4q}{(2\pi)^4} \left[
\frac{-4 \Delta_5}{3\Delta_1 \Delta_2^2}
+\frac{\frac{1}{3} \Delta_4 + 2r^2 \Delta_4 + 4r^4 \Delta_1^2}{\Delta_2^2} \right]
\nonumber \\
&& I_\Sigma = 4 \pi^2 \int_{-\pi}^{\pi} \frac{d^4q}{(2\pi)^4} \left[
\frac{\Delta_4}{4\Delta_1^3\Delta_2} (\Delta_3 - r^2 \Delta_1^2)
+ \frac{\Delta_5}{4\Delta_1^2\Delta_2} - \frac{r^2 \Delta_6}{\Delta_2}
-\frac{\Delta_1}{2} \right]
\nonumber
\eea
with
\bea
&&\Delta_1 = \sum_\mu \sin^2(q_\mu/2) \qquad;\qquad
\Delta_2 = \sum_\mu \sin^2(q_\mu) + 4r^2 \left[\sum_\mu \sin^2(q_\mu/2) \right]^2
\nonumber \\
&&\Delta_3 = \sum_\mu \sin^4(q_\mu/2) \qquad;\qquad
\Delta_4 = \sum_\mu \sin^2(q_\mu)
\\
&&\Delta_5 = \sum_\mu \sin^2(q_\mu) \sin^2(q_\mu/2) \qquad;\qquad
\Delta_6 = \sum_\mu \cos(q_\mu)
\nonumber
\eea
In the present work, the Wilson parameter has been set to unity; $r=1$.

The recipe for finding $q^\ast$ is most straightforward for the ratio
$Z_S/Z_P$. In order to implement eq.~(\ref{eq:scale}), we must simply
compute the 4-dimensional integral $I_2$ and also the corresponding
integral $L_2$, obtained by multiplying the integrand of $I_2$ by a $\ln(q^2)$.
$L_2$ is also finite, and therefore calculable. Thus, for $Z_S/Z_P$ we have
\be
\ln({q^\ast}^2) = \frac{L_2}{I_2}
\ee
We note in passing that the recipe of eq.~(\ref{eq:scale}) is not univocal,
even for this simple case: the internal momentum $q$ of the integral $I_2$,
being a dummy variable, can be chosen with a certain degree of arbitrariness
(in particular, translations of $q$) which does not affect the final answer,
since the integrand is made up of translationally invariant propagators.
However, such changes in $q$ will affect the logarithm in the integrand of
$L_2$, and thus will give us a different $q^\ast$. This is only an academic
problem, because $q^\ast$ is only meant to be a typical scale of the process
under examination. Thus, estimates of $q^\ast$ which differ by the choice
of internal momenta $q$ are all equally acceptable.

The calculation of $q^\ast$ for $Z_V$ and $Z_A$ is more complicated. Besides
$L_2$, we also need $L_1$ and $L_3$ (again obtained from $I_1$ and $I_3$,
by modifying their integrands by $\ln(q^2)$). $L_1$ and $L_3$ are also
finite, so this causes no serious problem. The real complication arises
from the unity on the r.h.s of eq.~(\ref{eq:Is}). This constant is obtained
as a difference of two integrals which diverge logarithmically with the
external momentum $ap^2$. We write them out explicitly. The first one arises
from the divergent part of the projected amputated Green functions
$\Gamma_V(p)$ and $\Gamma_A(p)$:
\bea
\hat I_\Gamma(ap)&=& 16\pi ^2\int_{-\pi }^\pi 
\dfrac{d^4q}{(2\pi )^4}\dfrac{\dsum_\rho \sin ^2q_\rho }{\left[ 4\dsum_\mu
\sin ^2\left( \dfrac{q_\mu }2\right) \right] ^2\left[ 4\dsum_\nu \sin
^2\left( \dfrac{q_\nu +ap_\nu }2\right) \right] } \nonumber \\
&=& -\ln (a^2p^2) - \gamma _E + F_{0001} + 5/2 + \calo(ap)
\eea
whereas the second integral arises from the divergent part of the self-energy:
\bea
\hat I_\Sigma (ap) &=& - 16\pi ^2\int_{-\pi }^\pi 
\dfrac{d^4q}{(2\pi )^4}\dfrac{\dsum_\rho \sin \left( q_\rho +ap_\rho
\right) \sin \left( q_\rho \right) }{\left[ 4\dsum_\mu \sin ^2\left( 
\dfrac{q_\mu }2\right) \right] ^2\left[ 4\dsum_\nu \sin ^2\left( \dfrac{%
q_\nu +ap_\nu }2\right) \right] } \nonumber \\
&=& \ln (a^2p^2) + \gamma _E - F_{0001} - 3/2 + \calo(ap)
\eea
The integration (in the limit $ap \rightarrow 0$) has been performed
with the aid of the tabulated integrals of \cite{ellis}. Thus, the factor
$+1$ on the r.h.s. of the first two eqs.~(\ref{eq:Is}) is obtained from the
integral $\hat I_{SUM} = \hat I_\Gamma + \hat I_\Sigma$:
\bea
\hat I_{SUM} &=& 16\pi ^2\int_{-\pi }^\pi
\dfrac{d^4q}{(2\pi )^4}\dfrac{\dsum_\rho
\left[ \sin \left( q_\rho \right)-\sin \left( q_\rho +ap_\rho \right)
\right] \sin \left( q_\rho \right) }{\left[ 4\dsum_\mu \sin ^2\left( 
\dfrac{q_\mu }2\right) \right] ^2\left[ 4\dsum_\nu \sin ^2\left( \dfrac{%
q_\nu +ap_\nu }2\right) \right] } \nonumber \\
&=& 1 + \calo(ap)
\label{eq:isum}
\eea
The above integral is finite. Its contribution to the RC's is given
by its limit $ap \rightarrow 0$.
We note that, in this limit the integrand of $\hat I_{SUM}$
vanishes, except for $q_\mu = 0$. Thus, the integral is dominated
by the region of small $q$. This enables us to substitute the integrand by
its small $q$ and $ap$ limit and take it over the whole $q$ range:
\be
\hat I_{SUM} = -\frac{1}{\pi^2} \int_{-\infty}^{+\infty} d^4q \frac{ap \cdot q}
{q^4 (q+ap)^2}
\ee
We introduce the standard Feynman parameters, and after some trivial
algebraic manipulations obtain:
\be
\hat I_{SUM} = - \frac{2}{\pi^2} \int_0^1 dx (1-x)
\int_{-\infty}^{+\infty} d^4q \frac{ap \cdot q}
{[(q+xap)^2 + (ap)^2 x (1-x)]^3}
\label{eq:isumx}
\ee
We next shift the integration variable $q \rightarrow q + xap$ and pass over
to polar coordinates in $q$-space. Integrating the polar angles leaves us
with
\be
\hat I_{SUM} = 2(ap)^2 \int_0^1 dx (1-x)
\int_0^{+\infty} dq^2 \frac{q^2}
{[q^2 + (ap)^2 x (1-x)]^3}
\label{eq:isumxx}
\ee
Finally, we perform a further change of variables $q^2 \rightarrow 
q^2/[a^2 p^2 x (1-x)]$, and carry out the integration in $x$ to obtain
\be
\hat I_{SUM} = 2 \int_0^{+\infty} dq^2 \frac{q^2}{(q^2 + 1)^3}
\label{eq:isumf}
\ee
The above is easily seen to be equal to 1, which is the final result of
eq.~(\ref{eq:isum}).

Now the application of eq.~(\ref{eq:scale}) requires the introduction of 
$\ln(q^2)$ in the integrand of $\hat I_{SUM}$.
However, it is not obvious at which stage of the successive integrations
from eq.~(\ref{eq:isum}) to eq.~(\ref{eq:isumf}) this should be done. For
example, if we were to introduce $\ln(q^2)$ in eq.~(\ref{eq:isumxx}), we would
find a result proportional to $\ln(a^2p^2)$. This does not make sense, since it 
yields a $q^\ast$ which depends on the momenta of the external states and, 
even worse, this $ap$ dependence is singular. Thus we prefer to introduce the 
$\ln(q^2)$ in the integrand of eq.~({\ref{eq:isumf}):
\be
L_{SUM} = 2 \int_0^{+\infty} dq^2 \frac{q^2 \ln(q^2)}{(q^2 + 1)^3}
\ee
The above integration can be easily performed; we find $L_{SUM} = 1$, which
is a $p$-independent constant. Thus, $q^\ast$ for $Z_V$ is given
by
\bea
\ln({q^\ast}^2) &=& \dfrac{1-\left[\dfrac{1}{4} L_1-\dfrac{1}{2}L_2-L_3 
+ L_\Sigma \right]}
{1-\left[\dfrac{1}{4} I_1-\dfrac{1}{2}I_2-I_3 + I_\Sigma \right]}
\eea
For $Z_A$ we have
\bea
\ln({q^\ast}^2) &=& \dfrac{1-\left[\dfrac{1}{4} L_1+\dfrac{1}{2}L_2-L_3 
+ L_\Sigma \right]}
{1-\left[\dfrac{1}{4} I_1+\dfrac{1}{2}I_2-I_3 + I_\Sigma \right]}
\eea

The Clover case presents no extra conceptual difficulties. Everything is
like in the Wilson case, except that on the r.h.s. of eqs.~(\ref{eq:Is}) we
have more finite integrals, which arise from the extra (Clover) term in the
action and the $\slashchar{D}$ field rotations of subsection \ref{sec:clover}.
These finite integrals are listed in refs.~\cite{gabri,ari}.

A final comment is in place here.
A shortcut to all the subtleties related to $L_{SUM}$ would be to simply
ignore the contribution of this integral to $q^\ast$. In this case, $q^\ast$
is obtained from the last two equations, with the $+1$'s omitted from
both numerator and denominator. With this $q^\ast$ the most affected
RC's are $Z_V$ and $Z_A$ of the Clover action. Their values change
by less than $2\%$, for all boosting recipes of the coupling.


\begin{thebibliography}{999}

\bibitem{msmith}
B. Meyer and C. Smith, Phys. Lett. B123 (1983) 62.

\bibitem{mz}
G.~Martinelli and Y.C.~Zhang, Phys.Lett. B123(1983) 433; 
Phys. Lett. B125 (1983) 77.

\bibitem{gaym}
A.~Gonzalez Arroyo, F.J.~Yndurain and G.~Martinelli, Phys. Lett. B117
(1982) 437.

\bibitem{groot}
R. Groot, J. Hoek and J. Smit Nucl Phys. B237 (1984) 111.

\bibitem{MARTIW}
G. Martinelli, Phys. Lett. B141 (1984) 395.

\bibitem{draper}
C. Bernard, A. Soni and T. Draper, Phys. Rev. D36 (1987) 3224.

\bibitem{gabri}
E.~Gabrielli et al., Nucl.Phys. B362 (1991) 475.

\bibitem{ari}
A.~Borrelli, R.~Frezzotti, E.~Gabrielli and C.~Pittori,
Nucl. Phys. B409 (1993) 382.

\bibitem{FREZZOTTI}
R. Frezzotti, E. Gabrielli, C. Pittori and G.C. Rossi, 
Nucl. Phys. B373 (1991) 781.

\bibitem{morn}
C.J. Morningstar, Nucl. Phys. B(Proc. Suppl.)47 (1996) 92.

\bibitem{mm}
L.~Maiani and G.~Martinelli, Phys.Lett. B178 (1986) 265.

\bibitem{guichris}
G. Martinelli and C.T. Sachrajda, Nucl. Phys. B306 (1988) 865;
Nucl. Phys. B316 (1989) 305.

\bibitem{sach}
C.T.~Sachrajda, Nucl. Phys. B(Proc.Suppl) 9 (1989) 121.

\bibitem{cris1}
M. Crisafulli, G. Martinelli, V.J. Hill and C.T. Sachrajda,
Phys. Lett. B223 (1989) 90.

\bibitem{masavla}
G.Martinelli, C.T.Sachrajda and A.Vladikas, Nucl. Phys.
B358 (1991) 212.

\bibitem{MARTIR}
G. Martinelli, C.T. Sachrajda, G. Salina and A. Vladikas, 
Nucl. Phys. B378 (1992) 591; Nucl. Phys. B397 (1993) 479.

\bibitem{clvwi}
G.~Martinelli, S.~Petrarca, C.T.~Sachrajda and A.~Vladikas,
Phys. Lett. B311 (1993) 241; Phys. Lett. B317 (1993) 660.

\bibitem{anti-KLM}
M. Crisafulli, V. Lubicz, G. Martinelli and A. Vladikas, 
Nucl. Phys. B(Proc. Suppl.)42 (1995) 400.

\bibitem{gri}
M.~Paciello, S.~Petrarca, B.~Taglienti and A.~Vladikas
Phys. Lett. B341 (1994) 187.

\bibitem{ukzs}
D.S. Henty, R.D. Kenway, B.J. Pendleton and J.I. Skullerud,
Phys. Rev D51 (1995) 5323.

\bibitem{jlqcd_lat96}
JLQCD Collaboration, S. Aoki {\em et al.}, 
Nucl.Phys.B(Proc.Suppl.) 53 (1997) 349;\\
see also preprint hep-lat/9705035.

\bibitem{np}
G.Martinelli et al, Nucl.Phys.B445(1995)81.

\bibitem{DS=2}
A. Donini et al, Phys. Lett. B360 (1996) 83.

\bibitem{schier}
M.~G\"ockeler et al., Nucl. Phys. B(Proc.Suppl) 47 (1996) 493.

\bibitem{vlad}
A. Vladikas, Nucl. Phys. B(Proc. Suppl.)47 (1996) 84.

\bibitem{wustl}
A. Donini, V. Gim\'enez, G. Martinelli, G.C. Rossi, M. Talevi, M. Testa 
and A. Vladikas, Nucl.Phys.B(Proc.Suppl.) 53 (1997) 883.

\bibitem{rossi}
G.C. Rossi, Nucl.Phys.B(Proc.Suppl.) 53 (1997) 3.

\bibitem{power}
G. Martinelli, Nucl. Phys. B(Proc. Suppl.)42 (1995) 127.

\bibitem{lep_mac}
G.P.~Lepage and P.B.~Mackenzie, Phys. Rev. D48 (1993) 2250.

\bibitem{sw}
B. Sheikholeslami and R. Wohlert, Nucl. Phys. B259 (1985) 572.

\bibitem{heatlie}
G.Heatlie et al, Nucl. Phys. B352 (1991) 266.

\bibitem{lusch1}
K.~Jansen et al.,Phys. Lett. B372 (1996) 275; \\
M. L\"uscher, S.~Sint, R.~Sommer and P.~Weisz, Nucl. Phys. 
B 478 (1996) 365; \\
M. L\"uscher et al., Nucl. Phys. B491 (1997) 323.

\bibitem{lusch}
M. L\"uscher, S.~Sint, R.~Sommer and H.~Wittig,
Nucl. Phys. B 491 (1997) 344.

\bibitem{mrsstt} G.~Martinelli, G.C.~Rossi, C.T.~Sachrajda, S.~Sharpe,
M.~Talevi and M.~Testa, Phys. Lett. B411 (1997) 141.

\bibitem{noi} C.~Dawson, V.~Lubicz, G.~Martinelli, G.C.~Rossi,
C.T.~Sachrajda, S.~Sharpe, M.~Talevi and M.~Testa, in preparation.

\bibitem{lep}
G.P.~Lepage, Nucl.Phys. B(Proc.Suppl.)26 (1992) 45.

\bibitem{kron}
A.S.~Kronfeld, Nucl.Phys. B(Proc.Suppl.)30 (1993) 445.

\bibitem{paul}
P.~B.~Mackenzie, Nucl.Phys. B(Proc.Suppl.)30 (1993) 35.

\bibitem{ekm}
A.X.~El-Khadra, A.S.~Kronfeld and P.B.~Mackenzie,
Phys. Rev. D55 (1997) 3933

\bibitem{wils1}
K.G.Wilson, Phys. Rev. D10 (1974) 2445.

\bibitem{wils2}
K.G.Wilson, in ``New Phenomena in Subnuclear Physics" (Erice 1975),
ed. A.Zichichi (New York, Plenum, 1975).

\bibitem{allmass}
C.R.Allton et al, Nucl.Phys.B431(1994)667. 

\bibitem {ciuc}
M. Ciuchini et al, Z.Phys. C68 (1995) 239.

\bibitem{brodski} S.~J.~Brodski, G.~P.~Lepage and P.~B.~Mackenzie,
Phys. Rev. D28 (1983) 228.

\bibitem{parisi} G.~Parisi, in ``High Energy Physics - 1980", Proceedings
of the XXth International Conference, Madison, Wisconsin, eds. L.~Durand
and L.~G.~Pondrom (American Institute of Physics, New York, 1981).

\bibitem{fbs62}
C.R.~Allton et al., Phys. Lett. B326 (1994) 295.

\bibitem{bknp}
M. Crisafulli et al., Phys. Lett. B369 (1996) 325.

\bibitem{ks}
L.H.~Karsten and J.~Smit, Nucl.Phys. B183 (1981) 103.

\bibitem{boc}
M.~Bochicchio et al., Nucl.Phys. B262 (1985) 331.

\bibitem{curci}
G.~Curci, Phys. Lett. B167 (1986) 265.

\bibitem{impr}
H.W. Hamber and C.M. Wu, Phys. Lett. B133 (1983)351;
Phys. Lett. B136 (1984)255;\\
W. Wetzel, Phys. Lett. B136 (1984) 407;\\
T. Eguchi and N. Kawamoto, Nucl. Phys. B237 (1984) 572.

\bibitem{sym}
K. Symanzik, in Mathematical Problems in Theoretical Physics,
Springer Lecture Notes vol.153 (1982) 47; eds. R. Schrader,
R. Seiler and D.A. Uhlenbrock;\\
K. Symanzik, Nucl. Phys. B226 (1983) 187;\\
K. Symanzik, Nucl. Phys. B226 (1983) 205.

\bibitem{naik}
S.~Naik, Phys. Lett. B311 (1993) 230.

\bibitem{bern}
C.W.~Bernard, Nucl.Phys. B(Proc.Suppl.)34 (1994) 47;
Phys.Lett. B125(1983) 77.

\bibitem{abada}
A.~Abada et.al. Nucl.Phys. B376(1992)172; \\
A.~Abada et.al. Nucl.Phys. B416(1994)675.

\bibitem{lw}
M.~L\"uscher and P.~Weisz, Nucl. Phys. B479 (1996) 429; \\
S.~Sint and P.~Weisz, Nucl.Phys. B502 (1997) 251.

\bibitem{guagn}
M. Guagnelli and R. Sommer, hep-lat/9709088.

\bibitem{ellis}
R.K.~Ellis and G.~Martinelli, Nucl. Phys. B235 (1984) 93.

\end{thebibliography}
\end{document}